\pdfoutput=1
\ifx\Reviewmode\undefined
\documentclass[journal,final,twoside]{IEEEtran}
\else
\documentclass[11pt,draftcls,onecolumn]{IEEEtran}
\fi
\usepackage{IEEEtran-fixes}
\usepackage[italian,english]{babel}
\usepackage{fixltx2e-fixes}
\usepackage[T1]{fontenc}
\usepackage[utf8x]{inputenc}
\usepackage{textcomp}
\usepackage[artemisia]{textgreek}
\usepackage[T1]{url}
\usepackage[unicode]{hyperref}
\usepackage[final]{microtype}
\usepackage{amsmath}
\usepackage[thmmarks, amsmath]{ntheorem}
\usepackage[romanConstants, boldVectors]{scstuff}
\usepackage{units}
\usepackage{xspace}
\usepackage{bm}
\usepackage{mdwtab}
\usepackage{acronym}
\usepackage{cite}
\usepackage[pdftex]{graphicx}
\usepackage[svgnames]{xcolor}
\usepackage{paralist}
\usepackage[caption=false]{subfig}
\usepackage[pdftex,xcolor]{changebar}
\usepackage{amsmath-matrix-ext}

\def\endIEEEproof{\hspace*{\fill}~\IEEEQED\par\vskip 0.5ex}

\hypersetup{colorlinks=true, 
  linkcolor=DarkBlue,
  citecolor=DarkBlue,
  urlcolor=DarkBlue}

\graphicspath{{./}{./Figures/}}

\makeatletter
\newcommand\isacroused[3]{%
  \expandafter\ifx\csname ac@#1\endcsname\AC@used
  #2\else #3\fi}
\makeatother
\newacro{FA}{Filtered Approximation}
\newacro{FIR}{Finite Impulse Response}
\newacro{IIR}{Infinite Impulse Response}
\newacro{STF}{Signal Transfer Function}
\newacro{NTF}{Noise Transfer Function}
\newacro{LP}{Low-Pass}
\newacro{HP}{High-Pass}
\newacro{BP}{Band-Pass}
\newacro{BS}{Band-Stop}
\newacro{SDP}{Semi-definite Programming}
\newacro{KYP}{Kalman–Yakubovich–Popov}
\newacro{A/D}{analog to digital}
\newacro{D/D}{digital to digital}
\newacro{D/A}{digital to analog}
\newacro{PDS}{Power Density Spectrum}
\newacroplural{PDS}{Power Density Spectra}
\newacro{PCM}{Pulse Code Modulation}
\newacro{PDF}{Probability Density Function}
\newacro{OSR}{Oversampling Ratio}
\newacro{LMI}{Linear Matrix Inequality}
\newacro{DCP}{Disciplined Convex Programming}

\newcommand\STF{\ensuremath{\mathit{STF}}}
\newcommand\NTF{\ensuremath{\mathit{NTF}}}
\newcommand\FF{\ensuremath{\mathit{FF}}}
\newcommand\FB{\ensuremath{\mathit{FB}}}
\newcommand\OSR{\ensuremath{\mathit{OSR}}}

\ifx\Reviewmode\undefined
  \newcommand\Titlesize\relax
\else
  \newcommand\Titlesize{\fontsize{22}{25}\selectfont}
\fi

\theorembodyfont{\itshape}
\theoremheaderfont{\normalfont\bfseries}
\theoremseparator{}

\newtheorem{property}{Property}

\theoremstyle{nonumberplain}
\theoremheaderfont{\normalfont\bfseries}
\theorembodyfont{\normalfont}
\theoremsymbol{~~\IEEEQEDopen}
\cbcolor{Red}

\newcounter{savecounterI}

\PrerenderUnicode{©}

\begin{document}
\ifx\Reviewmode\undefined\else
\newlength\hlw
\setlength\hlw{0.5\linewidth}
\def\lw{\hlw}
\fi
%
\title{\Titlesize Output Filter Aware Optimization of the Noise Shaping
  Properties of \texorpdfstring{ΔΣ}{Delta-Sigma} Modulators via Semi-Definite
  Programming}

\author{%
  \thanks{This is a post-print version of a paper published in the IEEE
    Transactions on Circuits and Systems --- Part I, Vol.~60, N.~9, pp.~2352 -
    2365, Sept.~2013. Available through DOI
    \href{http://http://dx.doi.org/10.1109/TCSI.2013.2239091}%
    {10.1109/TCSI.2013.2239091}. Always cite as the published version.
    \protect\\[1ex]
    Copyright © 2013 IEEE. Personal use of this material is permitted. However,
    permission to use this material for any other purposes must be obtained
    from the IEEE by sending a request to
    \url{pubs-permissions@ieee.org}.\protect\\[-1ex]}%
  Sergio~Callegari\extrainfo{,~\IEEEmembership{Senior~Member,~IEEE}} and %
  Federico~Bizzarri\extrainfo{,~\IEEEmembership{Member,~IEEE}}%
  \thanks{%
    S.~Callegari is with the Department of Electrical, Electronic, and
    Information Engineering "Guglielmo Marconi" (DEI) and the Advanced Research
    Center on Electronic Systems for Information and Communication Technologies
    "E. De Castro" (ARCES) at the University of Bologna, Cesena 47521,
    Italy. E-mail: \protect\url{sergio.callegari@unibo.it}.}%
  \thanks{%
    F.~Bizzarri is with the Dipartimento di Elettronica, Informazione e
    Bioingegneria, Politecnico di Milano, Italy. E-mail:
    \protect\url{bizzarri@elet.polimi.it}.}%
}

\initializeplaintitle[%
  \def\Titlesize{}
  \def\texorpdfstring#1#2{#2}]
\initializeplainauthor
\hypersetup{pdftitle=\plaintitle, 
  pdfauthor=\plainauthor, 
  pdfcreator={Sergio Callegari}}

\ifx\Reviewmode\undefined \markboth{IEEE Transactions on Circuits and Systems
  I}%
{Callegari \MakeLowercase{\textit{et al.}}: Output Filter Aware Optimization of
  the Noise Shaping Properties of ΔΣ Modulators\dots} \fi
%



\maketitle
\ifx\Reviewmode\undefined\else
\vspace{-1.2cm}%
\fi

\acused{D/D}
\acused{D/A}
\begin{abstract}
  The \ac{NTF} of ΔΣ modulators is typically designed after the features of the
  input signal. We suggest that in many applications, and notably those
  involving \ac{D/D} and \ac{D/A} conversion or actuation, the \ac{NTF} should
  instead be shaped after the properties of the output/reconstruction
  filter. To this aim, we propose a framework for optimal design based on the
  \ac{KYP} lemma and semi-definite programming. Some examples illustrate how in
  practical cases the proposed strategy can outperform more standard
  approaches.
\end{abstract}
%
\ifCLASSOPTIONpeerreview
\begin{center}
  \setcounter{page}{0}%
  \bfseries EDICS Category: ...
\end{center}
\fi

\IEEEpeerreviewmaketitle

\acresetall

\section{Introduction}
\label{sec:intro}
ΔΣ modulators \cite{Norsworthy:DSDC-1996, Reiss:JAES-56-1+2,
  Schreier:UDSDC-2004, DeLaRosa:TCAS1-58-1} are nowadays widely used in a
variety of systems, usually as \ac{A/D} and \ac{D/D} interfaces. The latter may
in turn simplify sample rate conversion, \ac{D/A} conversion, or power
amplification in actuation tasks. Typical applications range from data
conversion itself to signal processing in wide sense, including digital audio
\cite{Reefmann:SPDSD-2002}, frequency synthesis \cite{Yang:TVLSI-17-6},
switched mode amplification \cite{Gaalaas:AD-40-6, Ghannouchi:CASM-2010-4},
power conversion and actuation \cite{Dallago:TCAS1-44-8, Jacob:TIE-2012},
digital communications \cite{Galton:TMTT-50-1, Ghannouchi:CASM-2010-4}, sensing
\cite{Dong:SJ-7-1} and more. Recently, more exotic applications, such as in
optimization, have been proposed too \cite{Callegari:TSP-58-12,
  Bizzarri:ISCAS-2010, Bizzarri:CSDM-2010}.

ΔΣ modulators are signal encoders (or re-coders) capable of trading rate for
accuracy in order to translate high-resolution slowly (or non) sampled signals
into low-resolution rapidly sampled signals with little loss of fidelity. This
property is achieved through a feedback architecture involving a quantizer and
linear filters which provide \emph{noise shaping}, i.e., the ability to
unevenly distribute the quantization noise power so that some frequency bands
get most of it and others almost none.

\begin{figure}[t]
  \begin{center}
    \includegraphics[scale=0.52]{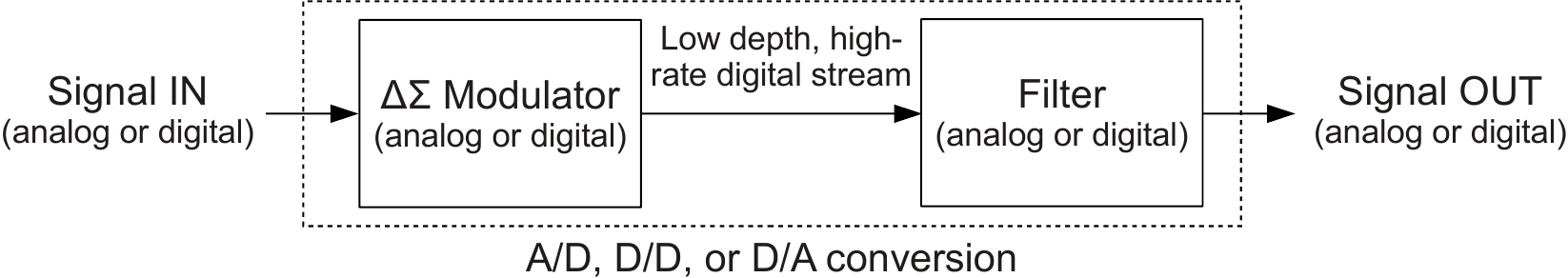}
  \end{center}
  \caption{Typical deployment of a ΔΣ modulator.}
  \label{fig:block-dia-generic}
\end{figure}

\begin{figure}[t]
  \begin{center}
    \includegraphics[width=\lw]{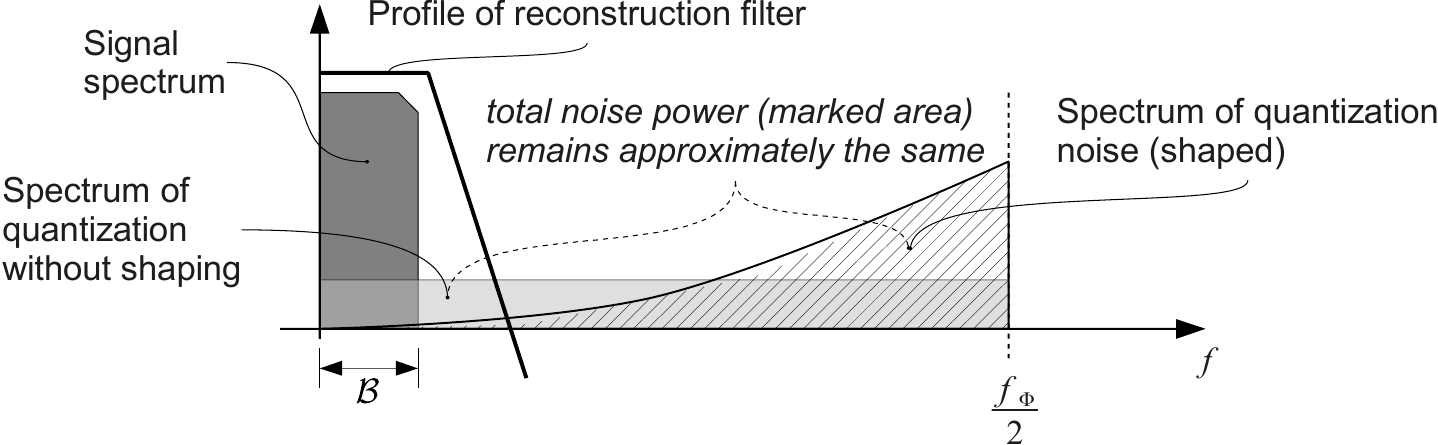}
  \end{center}\vskip -1ex
  \caption{Typical noise shaping and noise removal in a \ac{LP} ΔΣ modulator.}
  \label{fig:sample-spectra}
\end{figure}

\begin{figure}[t]
  \begin{center}
    \subfloat[\label{sfig:block-dia-adc}]{%
      \includegraphics[scale=0.52]{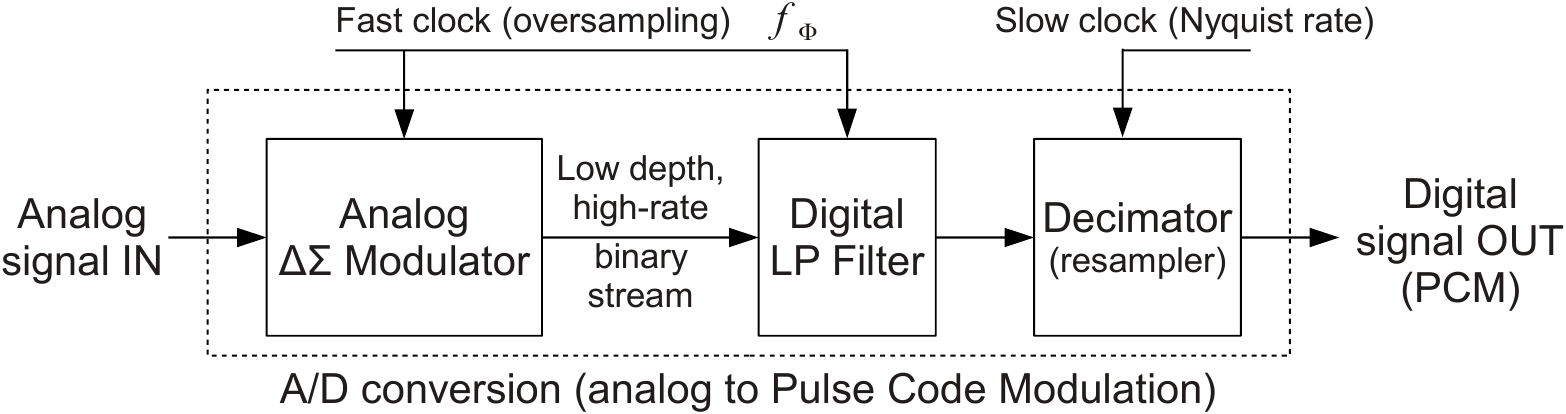}}\\
    \subfloat[\label{sfig:block-dia-dac}]{%
      \includegraphics[scale=0.52]{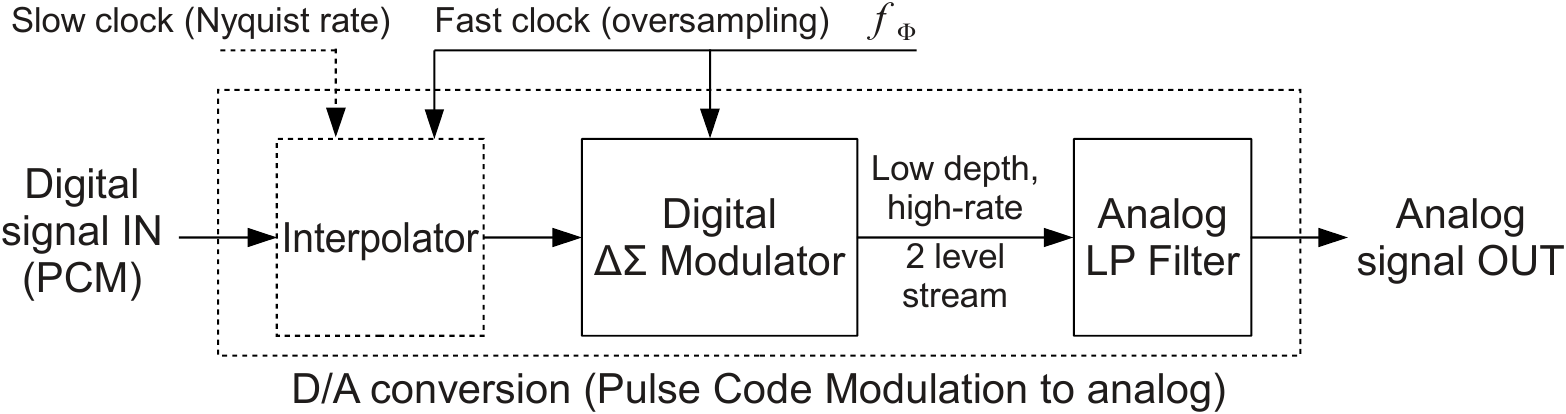}}\\
    \subfloat[\label{sfig:block-dia-ampli}]{%
      \includegraphics[scale=0.52]{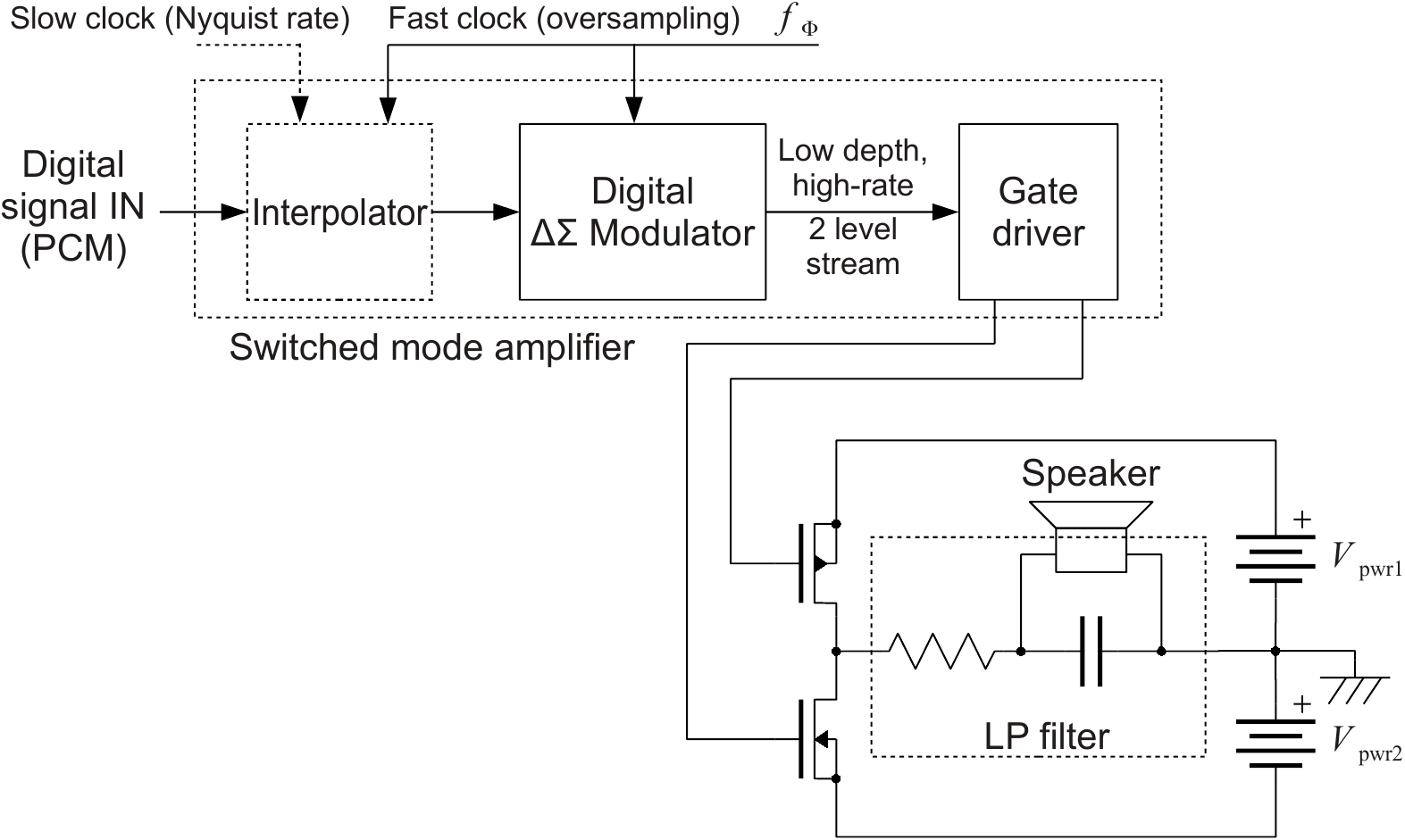}}
  \end{center}
  \caption{Specializations of the architecture in
    Fig.~\ref{fig:block-dia-generic} for \ac{LP} signals and a binary-output ΔΣ
    modulator: \ac{A/D} converter with \acs{PCM}
    output~\protect\subref{sfig:block-dia-adc}; \ac{D/A} converter with
    \acs{PCM} input~\protect\subref{sfig:block-dia-dac}; switched-mode
    amplifier with \acs{PCM} input~\protect\subref{sfig:block-dia-ampli}.}
  \label{fig:block-dia-specialized}
\end{figure}

ΔΣ modulators are almost invariably used in conjunction with filters as in
Fig.~\ref{fig:block-dia-generic}, to recover useful information that is
otherwise polluted by quantization noise.  In fact, the digital stream at the
output of the modulator has a much wider bandwidth than the input waveform,
thanks to its high sampling rate (oversampling). Furthermore, it contains two
components. The first one reflects the input signal itself, thus occupying just
the set of frequencies $\mathcal{B}$ constituting the signal band. The second
one is quantization noise, whose power is approximately fixed, depending on the
quantizer resolution. In principle, the noise \ac{PDS} extends throughout all
the available bandwidth. Actually, the modulator lets the noise \ac{PDS}
concentrate more in certain regions than in others. If these regions do not
overlap with $\mathcal{B}$, then a filter can be used to get rid of (most of
the) noise component without affecting the signal one. These considerations
make evident how an output or reconstruction filter is mandatory. Indeed, the
modulator role is precisely to \emph{shape the noise} so that it can be made
\emph{orthogonal} to the signal (and thus linearly separable).
Fig.~\ref{fig:sample-spectra} shows the typical behavior in the frequency
domain for a modulator suitable for \ac{LP} signals ($f_\Phi$ indicates
the modulator output rate). Fig.~\ref{fig:block-dia-specialized} shows
some specializations of the generic architecture to binary \ac{A/D} conversion,
\ac{D/A} conversion and switched-mode power amplification.

The above premises explain why typical design flows \cite{Schreier:UDSDC-2004,
  Schreier:DELSIG, Nagahara:TSP-60-6} want the modulator noise shaping
properties to be based only on the signal features (and notably the width of
$\mathcal{B}$). To be separable from the signal, the quantization noise needs
just to be as \emph{orthogonal} as possible to it. Hence, typical flows let the
modulator shape its noise \ac{PDS} so that it is as low as possible in the
signal band and (consequently) high elsewhere, with a transition between the
two regions as steep as possible to prevent superposition. Indeed, this is what
\emph{theoretically} enables the most thorough noise separation.

However, the fact that two items are \emph{theoretically} well separable does
not mean that they necessarily get well separated \emph{in practice}. Typical
design flows assure that a linear filter exists capable of guaranteeing an
almost perfect noise removal. Nevertheless, they cannot assure that such filter
is actually deployed. As a matter of fact, there are favourable situations
where the designer has very good control over the output filter. In this case,
conventional design flows are probably optimal. For instance, in \ac{A/D}
conversion (Fig.~\ref{sfig:block-dia-adc}) the output filter is digital so that
a good filter can be implemented without excessive cost. In other cases, the
designer has only a limited control over the output filter. For instance, in
\ac{D/A} conversion (Fig.~\ref{sfig:block-dia-dac}), one has an analog
reconstruction filter whose cost may rapidly grow with its specification (in
particular with roll-off). This situation may also arise in signal synthesis
\cite{Bizzarri:ECCTD-2009}. Even worse, there may be cases where the filter is
in part pre-assigned leaving the designer with extremely limited or no control
at all over it. For instance, in actuation (Fig.~\ref{sfig:block-dia-ampli})
the filter is often partially (if not completely) provided by the electric
machine used for the actuation itself. As an example, consider that the popular
Texas Instruments LM4670 switching audio amplifier is marketed as a
\emph{filterless} solution where the \ac{LP} filter is provided by the speaker
parasitic inductance and inertia (and possibly by the listener's ear)
\cite{TI:LM4670}. A similar situation may arise in ac motor drives
\cite{Bizzarri:ISCAS-2012, Callegari:ICECS-2012}.

\begin{figure}[t]
  \begin{center}
    \subfloat[\label{sfig:flow-traditional}]{%
      \includegraphics[scale=0.52]{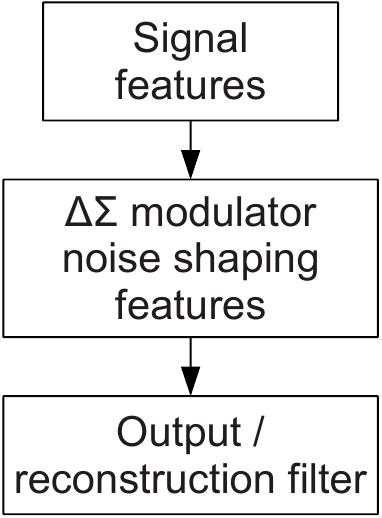}}\qquad
    \subfloat[\label{sfig:flow-new}]{%
      \includegraphics[scale=0.52]{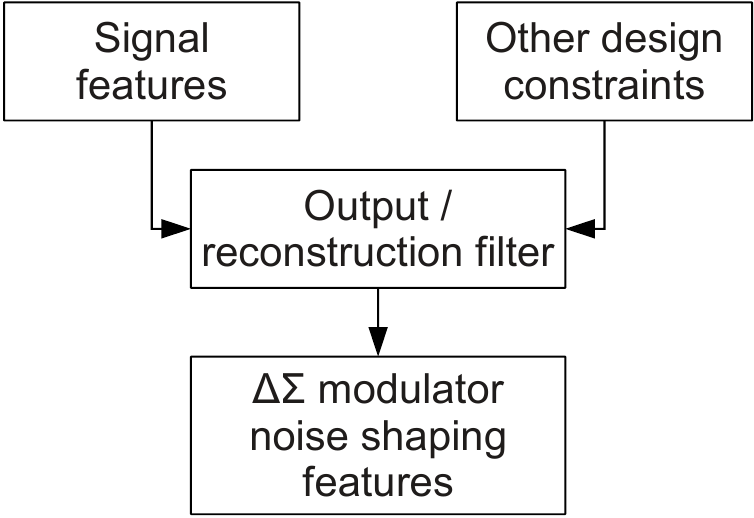}}
  \end{center}
  \caption{Traditional \protect\subref{sfig:flow-traditional} and proposed
    \protect\subref{sfig:flow-new} design flows.}
  \label{fig:flows}
\end{figure}

We claim that whenever the designer has limited or no control over the output
filter, the noise shaping properties of the ΔΣ modulator should not be designed
after the signal properties alone. Conversely, the designer, aware of the
limitations induced by a sub-optimal output filter, should explicitly consider
them to pursue the best possible reduction of the quantization noise. In the
following Sections, we formalize this claim providing a novel design flow for
ΔΣ modulators, based on the output filter features. Figure~\ref{fig:flows}
graphically summarizes the differences between a traditional flow and ours. In
the former \subref{sfig:flow-traditional}, the modulator noise shaping features
are designed after the signal properties alone. When these features are
obtained, an \emph{optimum} output filter is designed to take the best possible
advantage of them. In other words, the output filter is not a constraint, but a
degree of freedom to be exploited to make the most of the modulator noise
shaping profile. In our flow \subref{sfig:flow-new}, the first thing being
assigned is the output filter, for which not just the signal properties but
also other factors related to the context where the modulator is applied must
be considered (as it happens in the examples in Figs.~\ref{sfig:block-dia-dac}
and~\ref{sfig:block-dia-ampli}). Then, the modulator noise shaping features are
designed to cope at best with the filter. Thus, for us the output filter is a
constraint to be managed.

The proposed approach stems from interpreting the modulator as a heuristic
solver for a \ac{FA} problem \cite{Callegari:TSP-58-12}. It results in a
\ac{FIR} \ac{NTF}, obtained via \ac{SDP} \cite{Boyd:CO-2009}. The restriction
to \ac{FIR} \acp{NTF} often results in higher order modulators than in
conventional flows, but in many applications this is not an issue, as discussed
in \cite{Nagahara:TSP-60-6}. The minimization used to define the filter
coefficients respects the most important design constraints for ΔΣ modulators
\cite{Schreier:UDSDC-2004, Lee:Thesis-1987, Chao:TCAS-37-7} and thus enables a
robust design. It is formalized taking advantage of the \ac{KYP} lemma
\cite{Rantzer:SCL-28-1, Iwasaki:TAC-50-1}. This is not the first time that the
\ac{KYP} lemma is applied to the optimization of ΔΣ modulators, yet in
precedent cases the goals of the optimization were quite different
\cite{Nagahara:TSP-60-6, Osqui:ACC-2007}. Furthermore, previous attempts at
considering the output filter features in the design of ΔΣ modulators are
scarce and followed different strategies \cite{Gustavsson:TCAS1-57-12}.

In the last part of the paper, we provide extensive design examples, showing
how the proposed design strategy can consistently outperform conventional ones
and is also much more flexible, being capable of managing all kinds of
modulators, including multi-band ones, in a completely homogeneous
way. Furthermore, the strategy is often more robust. The examples show that, in
conjunction with output filters lacking too steep features, it results in
\acp{NTF} lacking steep features too. Consequently, as a positive side effect,
one often gets less extreme modulators that tend to be less prone to
misbehavior and deviation from expected performance.

\section{Notation}
For the sake of clarity and compactness, we make use of specific notations
relative to matrices and dynamical systems.

Matrices and vectors are generally indicated by capital italic letters in a
bold font as in $\mat A$, although for homogeneity with previous publications
and the Literature some vectors may be uncapitalized as in $\vec x$.  When it
is necessary to extract a sub-matrix from a matrix, the following notation
applies: $\mat A_{1:3,4:5}$ is the sub-matrix obtained from entries that belong
to the rows from 1 to 3 and the columns from 4 to 5 in $\mat A$. The colon is
saved if the two values on its sides are the same. For instance, $\mat
A_{1,4:5}$ is the sub-matrix (row-vector) obtained from values in the first row
and in columns from 4 to 5 in $\mat A$. A thick dot $\bullet$ can used as a
shorthand for \emph{beginning} or \emph{end} depending on the side of the colon
where it appears. For instance $\vec a_{2:\bullet}$ is a sub-vector containing
entries from the second to the last one in $\vec a$. Coherently with the
notation concepts illustrated above, $\bullet:\bullet$ can be replaced by a
single thick dot to be interpreted as a shorthand for \emph{all}. For instance,
$\mat A_{1,\bullet}$ is the first row of $\mat A$. The same notation can be
used to indicate matrix and vector elements. For instance, $\mat A_{2,1}$ is
the element at the second row, first column in matrix $\mat A$. Note that when
this kind of indexing is applied, indexes always start at 1.  In cases where
matrices or vectors need to be filled with values taken from other sequences,
parenthesis are used as in the following example: $\mat A = (a_{i,j})$ for $i
\in (0, \dots, 7)$ and $j \in (-1, \dots, 3)$ means that $\mat A$ is an
$8\times 5$ matrix, where $\mat A_{1,1} = a_{0,-1}$, and so on.  When it is not
otherwise declared, vectors are \emph{column} vectors. The transposition
operator $\transposed$ is often used to more compactly enumerate their entries
in a row, as in $\vec a = (a_1, a_2, a_3)\transposed$.

With respect to dynamical systems, a compact matrix notation is often used for
their state space model. For instance, if one has a discrete-time system
$\mathcal{G}$ with model
\[
\begin{cases}
  \vec x(k+1) = \mat A \vec x(k) + \mat B \vec u(k)\\
  \vec y(k) = \mat C \vec x(k) + \mat D \vec u(k)
\end{cases}
\]
one may write
\[
\mathcal{G}=
\left(\hspace{-0.5ex}
  \begin{array}{c|c}
    \mat A & \mat B\\[-0.5ex]
    \hlx{hv}
    \mat C & \mat D
  \end{array}
  \hspace{-0.5ex}\right)(z)
\]
where $(z)$ recalls the nature of the equations, based on time differences.

\section{Background}
\subsection{The ΔΣ modulator architecture and design constraints}
\label{ssec:design-constraints}
\begin{figure}[t]
  \begin{center}
    \subfloat[\label{sfig:ds-real}]{%
      \includegraphics[scale=0.55]{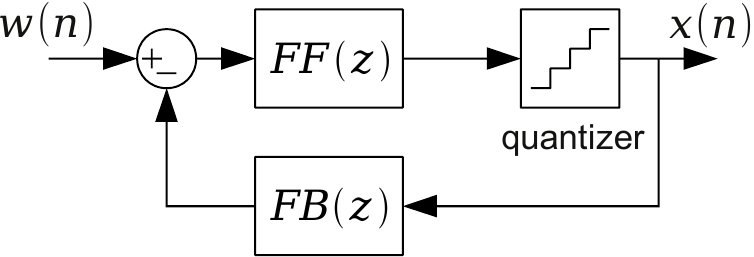}
    }\qquad
    \subfloat[\label{sfig:ds-linearised}]{%
      \includegraphics[scale=0.55]{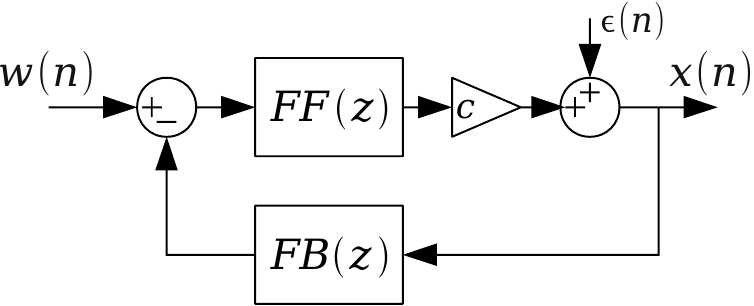}
    }%
  \end{center}
  \caption{General ΔΣ architecture and approximate linear model.}
  \label{fig:ds-generic}
\end{figure}

Fig.~\ref{sfig:ds-real} represents a generic architecture for a ΔΣ modulator
including a \emph{feedforward} filter $\FF(z)$, a \emph{feedback} filter
$\FB(z)$ and a quantizer.  All signals are assumed discrete-time and the
operation is timed by a fast clock with frequency $f_\Phi$.

Owing to the quantizer, the modulator is hard to tackle formally. A common
approach is to approximate it by the ``linearized'' architecture
in~\subref{sfig:ds-linearised}, where the quantisation noise $\epsilon(n)$ is
assumed to be white and independent of the input signal. The quantity $c$
models the \emph{average} quantizer gain. When the loop operates correctly
(namely, when the quantizer is not \emph{overloaded}), $c$ can be
assumed approximately unitary \cite{Schreier:UDSDC-2004, Nagahara:TSP-60-6} as
we actually take it in our discussion. In the same conditions, the quantization
noise amplitude can be assumed to be half the quantization step
$\Delta$. Typically, the quantization noise distributes approximately uniformly
in value, leading to a \ac{PDF} $\rho_\epsilon(x)$ approximately equal to
$\nicefrac{1}{\Delta}$ as long as $\nicefrac{-\Delta}{2} \le x \le
\nicefrac{\Delta}{2}$ and approximately null otherwise. This consideration
enables a first, rough estimation of the input noise power in the linearized
modulator, setting it at $\sigma^2_\epsilon = \int_{-\infty}^{\infty} x^2
\rho_\epsilon(x) \; d\!x = \nicefrac{\Delta^2}{12}$.  This is the average
quantization noise energy per sample. With it, the whiteness assumption on the
quantization noise lets one easily express the quantization noise \ac{PDS} over
the normalized angular frequency axis $\omega$ as
$E(\omega)=\nicefrac{\Delta^2}{12\pi}$. Note that we use single sided spectra
and $\omega\in[0,\pi]$.

The linearity of the approximate model is further exploited to decompose the
output in the contributions due to the input and to the quantisation noise
yielding $X(z) = \STF(z)\, W(z) + \NTF(z)\, E(z)$ where
$\STF(z)=\nicefrac{\FF(z)}{(1+ \FF(z)\, \FB(z))}$ and $\NTF(z)=\nicefrac{1}{(1+
  \FF(z)\, \FB(z))}$. Alternatively, if $\STF(z)$ and $\NTF(z)$ are assigned,
one has
\begin{equation}
  \begin{cases}
    \FF(z)=\frac{\STF(z)}{\NTF(z)}\\
    \FB(z)=\frac{1-\NTF(z)}{\STF(z)}.
  \end{cases}
  \label{eq:FF+FB}
\end{equation}
In typical applications, one wants the signal component to be passed from input
to output without alteration, so the \ac{STF} is unitary, and linear phase
(e.g, $\STF(z)=z^{-d}$, where $d$ is an integer delay).

It is now possible to discuss some constraints posed by the need to practically
implement the loop.
\begin{asparaenum}
\item For the stability of the linearized model, one needs the \ac{NTF} to be
  stable (the stability of the \ac{STF} is automatically guaranteed by taking
  it to be $z^{-d}$).
\item The \ac{NTF} needs to be causal (the causality of the \ac{STF} is
  automatically guaranteed by taking it to be $z^{-d}$).
\item $\FF(z)$ and $\FB(z)$ need to be causal.
\item The loop can not be algebraic. This means that $\FF(z) \FB(z) =
  \nicefrac{(1-\NTF(z))}{\NTF(z)}$, which represents the loop transfer
  function, needs to be strictly causal.
  \setcounter{savecounterI}{\theenumi}
\end{asparaenum}
Furthermore, it is necessary to observe that the conditions at point 1 are
necessary and sufficient for the stability of the linearized loop, but not
sufficient for the stability of the real loop, because the quantizer is
strongly non-linear. Determining strict conditions for the stability of the
non-linear loop is still an impossible task in general terms. However, the
analysis in \cite{Schreier:UDSDC-2004} and \cite[Sec.~IV-A]{Nagahara:TSP-60-6}
clearly indicate that such stability cannot be given just by the properties of
the loop filters, but must necessarily depend also on the maximum amplitude
that the input signal $w(n)$ takes over time, namely $\norm{w}_\infty =
\max_{n\in\Nset{N}} w(n)$, so that the stability is often more difficult to
achieve at large inputs. The same analysis ties the stability to the peak of
the amplitude response of the \ac{NTF}, indicating that the higher the peak
value, the more critical the stability. Following this consideration, practical
designs tend to rely on an empirical rule based on the limitation of the
\ac{NTF} gain (Lee criterion) \cite{Schreier:UDSDC-2004,
  Chao:TCAS-37-7}. Hence, one has a further constraint in addition to those
above:
\begin{asparaenum}
  \setcounter{enumi}{\thesavecounterI}
\item The peak of the \ac{NTF} amplitude response must be bounded to
  a low value, namely
  \begin{equation}
    \max_{\omega \in [0,\pi]} \abs{\NTF\left(\ee^{j\omega}\right)} < \gamma
    \label{eq:lee-constraint}
  \end{equation}
\end{asparaenum}
where the constant $\gamma$ depends on the number of quantization levels. For
binary quantizers, $\gamma$ should be less than $2$ and is typically set at
$1.5$. Incidentally, note that when a modulator turns out to overload its
quantizer, it is often possible to retry its design with a lower $\gamma$. For
modulators where the \ac{NTF} is high-order, which are more subject to
misbehavior, it is frequent to reduce $\gamma$ to $1.4$ or even slightly lower
values. However, having to reduce $\gamma$ too much also reduces the
effectiveness of the \ac{NTF}.

From the discussion about point 5, it should be clear that differently from the
previous four, this is neither a necessary nor sufficient condition. Rather, it
is just a requirement capable of making the ΔΣ modulator more \emph{likely} to
operate correctly in a wide range of practical cases.

Before proceeding to introduce design flows, it is worth mentioning that it can
be convenient to slightly reword the criteria 1-4. With reference to point 4,
say that the \ac{NTF} is $\nicefrac{B_\NTF(z)}{A_\NTF(z)}$. The loop function
is thus $\nicefrac{(A_\NTF(z)-B_\NTF(z))}{A_\NTF(z)}$. To have it strictly
causal, the order of $A_\NTF(z)$ must be the same as the order of
$B_\NTF(z)$. Hence, one has
\begin{asparaenum}
\item[2a)] The \ac{NTF} needs to be causal but not strictly causal.
\end{asparaenum}
Furthermore, there must be cancellations (at least one) in
$A_\NTF(z)-B_\NTF(z)$. The first cancellation can only happen if
\begin{asparaenum}
\item[4a)] The first coefficient in the impulse response of the \ac{NTF} is
  unitary.
\end{asparaenum}
If 2a is satisfied, the causality of $\FF(z)$ is always guaranteed, while the
causality of $\FB(z)$ can certainly be guaranteed by taking $\STF(z)=1$. Hence,
one can, with no loss of generality, always take $\STF(z)=1$ and consider
conditions 1, 2a, 4a, 5 instead of 1-5. This makes the modulator design
completely determined after the design of the \ac{NTF}.

\subsection{Conventional design flows}
From the Introduction, it should be clear that common design flows place great
attention to the noise present at the output of the modulator in the signal
band $\set{B}$. Note that $\set{B}$ is a subset of the normalized angular
frequency interval $[0,\pi]$ that may contain multiple sub-bands. The in-band
noise power at the output of the modulator is
\begin{multline}
  \sigma^2_{\set{B}} = \int_{\set{B}} 
  E(\omega) \abs{\NTF\left(\ee^{\ii\omega}\right)}^2 \, d\omega = \\
  \frac{\Delta^2}{12\pi} \int_{\set{B}} 
  \abs{\NTF\left(\ee^{\ii\omega}\right)}^2 \, d\omega .
  \label{eq:inbandnoise}
\end{multline}

For instance, in the renown case of a first order, binary \ac{LP} modulator
with $\FF(z)=\nicefrac{1}{z-1}$ (feedforward path is an accumulator) and
$\FB(z)=1$, one has $\STF(z)=z^{-1}$ and $\NTF(z)=1-z^{-1}$. The \ac{NTF} is a
first order differentiator having the \ac{HP} response
\begin{equation}
  \abs{\NTF\left(\ee^{\ii\omega}\right)}=
  \abs{1-\ee^{-\ii\omega}}=
  2 \sin(\nicefrac{\omega}{2}) .
  \label{eq:ntf-response-1}
\end{equation}
Letting $B$ be the overall width of set $\mathcal{B}$ and defining the \ac{OSR}
as $\OSR=\nicefrac{f_\Phi}{2B}$, in this \ac{LP} case the integral in
Eqn.~\eqref{eq:inbandnoise} turns out to be computed on
$[0,\nicefrac{\pi}{\OSR}]$ and thus reduces to the well known result
\begin{equation}
  \sigma^2_{\mathcal{B}} = \frac{\Delta^2}{12\pi} 
  \int_{0}^{\nicefrac{\pi}{\OSR}} 4 \sin^2(\nicefrac{\omega}{2})\,
  d\omega \approx \frac{\Delta^2}{12} \frac{\pi^2}{3 \OSR^3}
  \label{eq:noise-diff-1}
\end{equation}%
where the approximation is valid when the \ac{OSR} is large enough.  The result
above can be generalized to higher order \acp{NTF} taking
\begin{equation}
  \NTF(z)=(1-z^{-1})^P=\frac{(z-1)^P}{z^P}
  \label{eq:ntf-differentiator-P}
\end{equation}
i.e., making the \ac{NTF} a $P$ order differentiator (all poles in $0$ and all
zeros in $1$). This changes Eqn.~\eqref{eq:noise-diff-1} into the more general

\begin{equation}
  \sigma^2_{\mathcal{B}} \approx \frac{\Delta^2}{12} \frac{\pi^{2P}}{(2P+1)
    \OSR^{2P+1}} .
  \label{eq:noise-diff-P}
\end{equation}%
Three things are worth noticing: (i) the \ac{NTF} in
Eqn.~\eqref{eq:ntf-differentiator-P} is only suitable for \ac{LP} modulators;
(ii) it does not respect criterion 5 (for instance the amplitude response in
Eqn.~\eqref{eq:ntf-response-1} peaks at 4); and (iii) it does not minimize
$\sigma^2_{\set{B}}$.

Conventional design flows consequently aim at choosing a $P$ order \ac{NTF}
minimizing the expression in \eqref{eq:inbandnoise} while respecting
requirements 1, 2a, 4a, 5. Such a minimization is by no means easy and is
generally practiced with approximation and iterative methods. The Literature
proposes different variants. Notable ones are thoroughly described in
\cite{Schreier:UDSDC-2004} and effectively implemented in
\cite{Schreier:DELSIG}. An alternative strategy still based on the signal
features alone aims at a \emph{min-max} optimization of the quantization noise,
i.e., at minimizing the peak of the integrand in \eqref{eq:inbandnoise}, rather
than the integral itself\cite{Nagahara:TSP-60-6}.

Often, a key idea is to initially focus on a \ac{LP} modulator (namely, a
\ac{HP} \ac{NTF}), which can later be mutated into a \ac{BP} modulator if
needed (namely, transforming the \ac{NTF} into a \ac{BS} transfer function).
Intuitively, a starting point can be Eqn.~\eqref{eq:ntf-differentiator-P},
which satisfies requirement 4a.  The zeros can then be moved away from $z=1$
(dc) to spread them onto the portion of the unit circle corresponding to
frequency values from $0$ to $\nicefrac{\pi}{\OSR}$. This guarantees that the
\ac{NTF} can remain more uniformly low in the signal band. An optimal zero
placement can be obtained by considering \eqref{eq:inbandnoise} for an \ac{NTF}
with $P$ zeros placed onto the unit circle and by nulling its gradient taken
with respect to such zeros. In \cite{Schreier:UDSDC-2004} optimal values are
tabled for $P=1,\dots, 8$. A second step is to push the poles away from $0$
closer to $z=1$, letting them lay within the unit circle onto a curve
surrounding $z=1$ and confining the portion of the unit circle corresponding to
the signal bandwidth. This has the effect of limiting the gain of the \ac{NTF}
out of the signal band and is instrumental in respecting requirement 5. Some
common assumptions used in this optimization are that: (i) the zeros of the
\ac{NTF} can be assigned (almost) independently of the poles (namely, the poles
have negligible effect on the in-band noise or, alternatively, the denominator
of the \ac{NTF} has an almost flat response in the signal band); and (ii)
requirement 5 can be verified by assuming that the \ac{NTF} peaks at
$\omega=\pi$.

As an example, the approach above is implemented in the \texttt{synthesizeNTF}
function in the well known DELSIG toolbox \cite{Schreier:DELSIG}. The idea is
to take the \ac{NTF} zeros in $z=1$ or to spread them according to the
minimization procedure described above and then to take the poles of the
\ac{NTF} so that they correspond to a maximally flat \ac{LP} filter. The
bandwidth of the filter implied by the poles is then adjusted until requirement
5 is satisfied. In practice an independent optimization of the zeros and the
poles of the \ac{NTF} is practiced, which may lead to some issues for
particular design specifications \cite[Sec.~8.1]{Schreier:UDSDC-2004}.

A slight generalization of this procedure consists in choosing an \ac{NTF}
approximation type (e.g., Butterworth, inverse Chebyshev, etc.), and designing
a \ac{HP} \ac{NTF} so that it is in the form $\prod_{i=1}^{P}
\nicefrac{(z-z_i)}{(z-p_i)}$, where $z_i$ and $p_i$ are the zeros and poles (so
as to satisfy requirements 2a and 4a), and it has a cut-off angular frequency
$\omega_t$ \cite[Sec.~4.4.1]{Norsworthy:DSDC-1996}. Initially $\omega_t$ is set
just slightly above the upper edge of the signal bandwidth. Then, the value of
$\abs{\NTF(-1)}$ is verified and the filter is iteratively re-designed with
different values of $\omega_t$, until condition 5 is fulfilled. This means
reducing $\omega_t$ if the \ac{NTF} peaks at too high values and enlarging it
otherwise. For some filter forms, alternatively to (or together with)
$\omega_t$, the stop-band gain (or ripple) can be used as a degree of freedom
to satisfy condition 5. For instance, the DELSIG
\texttt{synthesizeChebyshevNTF} uses an inverse Chebyshev form (i.e., Chebyshev
filter with ripple in the stopband) \cite{Schreier:DELSIG}.

As a final remark, note that the most recent design flows may rely on more
sophisticated optimization strategies \cite{Nagahara:TSP-60-6}, but still base
them on the signal features (namely, on $\mathcal{B}$) alone.%

\subsection{Criticism of conventional design flows}
\label{sec:issues}
Here we briefly summarize some potential issues with conventional design flows.
\begin{enumerate}[a)]
\item Conventional flows share a major trait in assuming that a modulator would
  be \emph{perfect} if it could push all the quantization noise away from the
  signal band. However, ΔΣ modulators are almost always used together with
  output/reconstructions filters as in Figs.~\ref{fig:block-dia-generic}
  and~\ref{fig:block-dia-specialized}. Thus, this view of perfection in the
  modulator implies another assumption of perfection on the filter side. For a
  perfect modulator, the modulator+filter ensemble can work optimally only if
  the filter can let through all that is in the signal band and reject all that
  is outside. Unfortunately, this on-off behavior is impossible to
  implement. When the output filter is imperfect, a modulator that is perfect
  under this standard can lead to an overall modulator+filter behavior worse
  than that of an imperfect modulator.  Thus, in many cases one is better off
  adopting a different view of perfection, taking into account the features of
  the output filter from the very start. This is particularly important in
  cases like those in Figs.~\ref{sfig:block-dia-dac}
  and~\ref{sfig:block-dia-ampli} where the output filter is analog and its
  specifications cannot be made too strict without incurring into high
  costs. Nonetheless, even in cases like Fig.~\ref{sfig:block-dia-adc}, where
  the output filter is digital, taking its response into account can be
  beneficial. In fact, the filter is functional to decimation and the
  out-of-band noise that may leak out of it is no less important than the
  in-band noise since the decimator/resampler aliases it onto the signal band.

\item Many conventional design flows, where the location of the \ac{NTF} zeros
  is decided independently from the \ac{NTF} poles, provide good results only
  if the denominator polynomial of the \ac{NTF} turns out to be almost constant
  in the signal band. In some cases, and particularly when the \ac{OSR} is
  relatively low, this assumption may prove untrue, leading to a sub-optimal
  design.

\item In some design flows, the cut-off frequency of the \ac{NTF} is used as a
  degree of freedom to satisfy the constraint $\norm{\NTF}_\infty<\gamma$. In
  some cases, particularly when $\gamma$ is close to $1$ or the \ac{OSR} is
  low, this may result in some noise leaking into the signal band.

\item Most conventional design flows result in \acp{NTF} with steep transitions
  between the pass band and the stop band, particularly when the \ac{NTF} is
  high order. This behavior is obviously inherent in the minimization of
  $\sigma^2_{\set{B}}$. However, it may exacerbate the differences between the
  linearized and real modulator model. In turn, this may lead to inaccurate SNR
  predictions and in extreme cases to a lower robustness against
  instability.\footnote{%
    Even if we cannot provide a formal proof of this phenomenon, we have
    observed it in many of cases and it has intuitive explanations. We report
    one. If a modulator could be implemented \emph{fully respecting} the
    specification of an extremely steep NTF (e.g., brick-wall), then violations
    of information theory principles could occur. In fact, one could recover
    the modulator input information \emph{without any loss due to quantization}
    by using a brick-wall output filter, thus obtaining an information rate at
    the modulator output equal to that at the input. Since the latter can be
    arbitrarily large, this could imply an output information rate higher than
    the bit-rate, which is absurd. Thus, one can expect the conformance between
    the approximated linear model and the actual nonlinear model to deteriorate
    as the modulator is designed to have steeper NTFs, bringing in unexpected
    effects potentially including instability.}

\item Many conventional design flows start with an \ac{LP} modulator and obtain
  other modulator types (e.g. \ac{BP}) via transformations. This makes it
  extremely hard if not impossible to cope with unusual modulator types (e.g.,
  multi-band) that may be required by some applications.
\end{enumerate}

\section{The proposed \ac{NTF} optimization}
In this Section, we mainly deal with point a) in the list above. Nonetheless,
as a side effect, the proposed solution also addresses all the other points.

It is worth anticipating that our design strategy assumes and requires the
\ac{NTF} to be \ac{FIR}. In practice, this is not a severe limitation. As a
matter of fact, this choice is perfectly in line with the elementary, original
high-order \ac{NTF} form in Eqn.~\eqref{eq:ntf-differentiator-P}, where all the
poles fall in the origin. With respect to this, we merely move the
zeros. Consequently, our proposal can be seen as a strategy where only the
\ac{NTF} zeros are optimized. Lack of optimization for the poles means that
results comparable to strategies which optimize the poles can only be achieved
at a higher filter order. Indeed, this is the case. Yet, taking a higher filter
order is not a problem since contrarily to conventional designs, we can
synthesize high order \acp{NTF} (even 30-50 or more) without hindering the
modulator stability. Furthermore, other \ac{FIR} based strategies exist
\cite{Nagahara:TSP-60-6}, also requiring large modulator orders.

\subsection{Form to be optimized}
\begin{figure}[t]
  \begin{center}
    \includegraphics[scale=0.55]{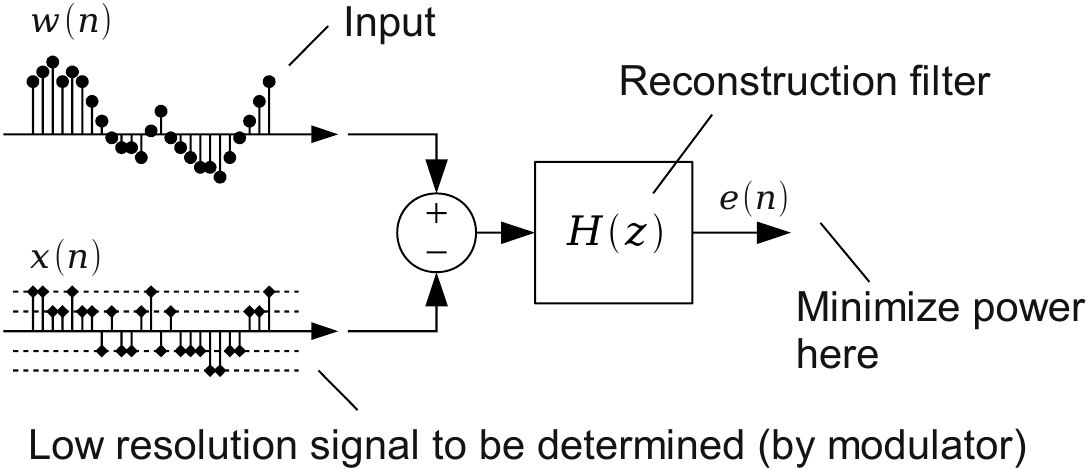}
  \end{center}
  \caption{Interpretation of the modulator operation as the solution of an
    \ac{FA} problem.}
  \label{fig:filtered-approx}
\end{figure}

As hinted in \cite{Callegari:TSP-58-12,Bizzarri:ECCTD-2009}, in order to deal
with the output filter, one should interpret the ΔΣ modulator as a heuristic
solver for an \ac{FA} problem. Fig.~\ref{fig:filtered-approx} illustrates the
problem nature. This consists in finding a discrete sequence $x(n)$ such that
it is as similar as possible to a high-resolution or continuous valued input
sequence $w(n)$, once passed through a filter $H(z)$. Clearly, $w(n)$ plays the
role of the modulator input, $x(n)$ of the modulator output and $H(z)$ is the
output filter. As shown in the figure, the concept of ``similarity'' can be
formalized as a minimization of the average power at the output of the filter
fed by $w(n)-x(n)$.

To solve the \ac{FA} problem, the ΔΣ modulator rather than being designed after
the minimization of $\sigma^2_{\set{B}}$ in Eqn.~\eqref{eq:inbandnoise} needs
to be designed after the minimization of
\begin{multline}
  \sigma^2_{H} = \int_{0}^\pi E(\omega)
  \abs{\NTF\left(\ee^{\ii\omega}\right)}^2 \,
  \abs{H \left(\ee^{\ii\omega}\right)}^2 \, d\omega = \\
  \frac{\Delta^2}{12\pi} \int_{0}^{\pi}
  \abs{\NTF\left(\ee^{\ii\omega}\right)}^2 \, \abs{H
    \left(\ee^{\ii\omega}\right)}^2\, d\omega .
  \label{eq:filterednoise}
\end{multline}

Let us assume that $\NTF(z)$ is achieved by a $P$ order FIR filter, with
coefficients $a_i$, collectable in a vector $\vec a = (a_0, \dots,
a_{P})\transposed$. Namely,
\begin{equation}
  \NTF(z)=\sum_{k=0}^{P} a_k z^{-k} .
\end{equation}
Let us also assume that $H(z)$ corresponds to a filter whose impulse response
can safely be truncated to a finite number of samples collectable in a vector
$\vec h=(h_0,\dots,h_{M})\transposed$.
Let us finally define $G(z)=\NTF(z) H(z)$, so that the quantity object of
minimization can be written as $\int_{0}^{\pi} \abs{G(\ee^{\ii\omega})}^2
\:d\!\omega$. The impulse response corresponding to $G(z)$ can obviously be
obtained as the convolution of $a_i$ and $h_i$, as in
\begin{equation}
  g_i=[a * h]_i = \sum_{j=-\infty}^{\infty} a_j\, h_{i-j}
\end{equation}
where $*$ is the convolution operator and we take $a_i=0$ for $i<0$ or $i>P$
and $h_i=0$ for $i<0$ or $i>M$. Clearly, there are only a finite number of
non-null entries in $g_i$, for $i=0,\dots,M+P$.

Eventually, recall the discrete form of Parseval's theorem, referred to $G(z)$
\begin{equation}
  \frac{1}{2\pi}\int_{-\pi}^{\pi} \abs{G(\ee^{\ii\omega})}^2 \:d\!\omega =
  \sum_{i=-\infty}^{\infty} |g_i|^2 .
\end{equation}

By substitution, neglecting all multiplicative constant terms, the quantity
object of minimization can be expressed as
\begin{equation}
  \sum_{i=0}^{M+P} \left( \sum_{j=0}^{P} a_j\, h_{i-j} \right)^2
\end{equation}
that can be further expanded into
\begin{equation}
  \sum_{i=0}^{M+P}\; \sum_{j=0}^{P}\; \sum_{k=0}^{P}
  a_j\, a_k\, h_{i-j}\, h_{i-k} .
\end{equation}
By swapping the sums, one gets
\begin{equation}
  \sum_{j=0}^{P}\; \sum_{k=0}^{P}
  a_j\, \left(\sum_{i=0}^{M+P}h_{i-j}\, h_{i-k}\right)\; a_k  .
\end{equation}%
that can be put in a more compact form exploiting a $(P+1)\times (P+1)$ matrix
$\mat Q$ defined as
\begin{equation}
  \mat Q = (q_{j,k}) ~~ \text{where} ~~
  q_{j,k} = \sum_{i=0}^{M+P}  h_{i-j}\, h_{i-k}
\end{equation}
for $j, k \in \{0,\dots, P\}$. With this, the form to be minimized becomes
\begin{equation}
  \vec a\transposed\; \mat Q\; \vec a .
  \label{eq:quadratic-minimization}
\end{equation}
Hence it is evident that one has a \emph{quadratic optimization problem}
defined over the filter coefficients. Clearly, $\mat Q$ must be positive
semi-definite, since $\sigma^2_H$ in~\eqref{eq:filterednoise} cannot be
negative.

It is worth observing that in the definition of $\mat Q$ the summation can be
extended to infinity as in
\begin{equation}
 q_{j,k} = \sum_{i=-\infty}^{\infty}  h_{i-j}\, h_{i-k} 
\end{equation}
to make evident that $\mat Q$ is Toeplitz \cite{Gray:FTCIT-2-3} symmetric (it
is in fact an auto-covariance matrix). This is interesting not just as a
structural property, but also to reduce the computational burden of $\mat Q$ to
the mere computation of its first row or first column. Focusing on the first
row $q_{0,k} = \sum_{i=-\infty}^{\infty} h_{i}\, h_{i-k}$ one can also notice
that to further reduce the computation burden the summation bounds can be
restricted to $q_{0,k} = \sum_{i=k}^{M} h_{i}\, h_{i-k}$.

\subsection{Application of the design constraints}
The particular choice of a \ac{FIR} structure for the \ac{NTF} makes its
stability (point 1 in Sec.~\ref{ssec:design-constraints}) and causality (point
2a) inherent. Thus, only two constraints remain: the need for a unitary first
coefficient in the impulse response (point 4a); and the containment of
$\norm{\NTF}_\infty$ (point 5, Lee criterion).  \smallskip

\subsubsection{Unitary the first coefficient of the impulse
  response}
thanks to the \ac{FIR} nature of the \ac{NTF} this merely requires fixing $\vec
a_1 = a_0 = 1$. While this equality can be taken as a constraint for the
minimization of \eqref{eq:quadratic-minimization}, it is actually more
convenient to use it to reduce the problem size. To this aim, observe that

\begin{multline}
  \vec a\transposed\;\mat Q\;\vec a =\\
  \begin{pmatrix}
    \vec a_1 & \vec a_{2:\bullet}\transposed
  \end{pmatrix}\;
  \begin{pmatrix}
    \mat Q_{1,1} & \vec Q_{1,2:\bullet}\\
    \vec Q_{2:\bullet,1} & \mat Q_{2:\bullet,2:\bullet}
  \end{pmatrix}\;
  \begin{pmatrix}
    \vec a_1\\\vec a_{2:\bullet}
  \end{pmatrix}=
  a_0^2\, q_{0,0} +\\
  a_0\, \vec a_{2:\bullet}\transposed\;\vec Q_{2:\bullet,1}+
  a_0\, \vec Q_{1,2:\bullet}\; \vec a_{2:\bullet}+
  \vec a_{2:\bullet}\transposed\;\mat Q_{2:\bullet,2:\bullet}\;\vec
  a_{2:\bullet} .
\end{multline}%

Thanks to the symmetry of $\mat Q$, the two central entries can be rewritten as
$2 a_0\; \mat Q_{1,2:\bullet}\; \vec a_{2:\bullet}$.  After the constantness of
$q_{0,0}$ and thanks to $a_0=1$ the quantity object of minimization eventually
reduces to
\begin{equation}
  \vec a_{2:\bullet}\transposed\;
  \mat Q_{2:\bullet,2:\bullet}\;
  \vec a_{2:\bullet} + 
  2\; \mat Q_{1,2:\bullet}\; \vec a_{2:\bullet} .
  \label{eq:objective}
\end{equation}
In other words, the constraint can be exploited for the reduction of the
problem size at the mere cost of augmenting the minimization form by a linear
term.

\subsubsection{Lee criterion}  
Eqn.~\eqref{eq:lee-constraint} represents an extremely complicated constraint.
Indeed, it can be recast based on a \emph{universal quantification}
\begin{equation}
  \forall \omega \in [0,\pi] \quad \abs{\NTF\left(\ee^{\ii\omega}\right)} 
  < \gamma
\end{equation}
making evident that it summarizes an \emph{infinitely large} set of
inequalities in the frequency domain. Furthermore, the filter coefficients
appear in $\abs{\NTF\left(\ee^{\ii\omega}\right)}$ in a nonlinear fashion.

Fortunately, the \ac{KYP} lemma provides an extremely efficient way to convert
universally quantified frequency domain inequalities of this sort into an
alternative formulation based on the dual \emph{existential quantifier}. This
is extremely convenient since a minimization problem can often deal with
existential quantifiers with the mere introduction of dummy
variables. Furthermore, the \ac{KYP} lemma gets the frequency domain inequality
expressed through an arbitrary realization of a dynamical system providing the
frequency domain behavior. This is also very convenient since it means that an
inequality over the \ac{NTF} can be directly expressed via the \ac{NTF}
coefficients.

From the \ac{KYP} lemma, the following property holds.

\begin{property}[Bounded real lemma]
  \label{prop:bounded-real}\hnull\\
  If a transfer function $T(z)$ admits a state space representation
  $\mathcal{T}$ such that
  \begin{equation}
    \mathcal{T}=
    \left(\hspace{-0.5ex}
      \begin{array}{c|c}
        \mat A & \mat B\\[-0.5ex]
        \hlx{hv}
        \mat C & \mat D
      \end{array}
      \hspace{-0.5ex}\right)(z)
  \end{equation}
  and $\mathcal{T}$ is stable and controllable, then an inequality such as
  \begin{equation}
    \norm{T}_\infty \leq \gamma
  \end{equation}
  can be recast in terms of the coefficients in $\mat A$, $\mat B$, $\mat C$
  and $\mat D$ by asserting that
  \begin{multline}
    \exists \mat P 
    \text{ square symmetric positive definite matrix such that}\\
    \begin{pmatrix}
      \mat A\transposed\,\mat P\,\mat A - \mat P &
      \mat A\transposed\,\mat P\,\mat B &
      \mat C\transposed\\
      \mat B\transposed\,\mat P\,\mat A &
      \mat B\transposed\,\mat P\,\mat B - \gamma^2&
      \mat D\\
      \mat C & \mat D & -1
    \end{pmatrix} \le 0
    \label{eq:matrix-inequality}
  \end{multline}
  where the $\le$ sign is here used to denote a generalized inequality stating
  negative semi-definiteness.
\end{property}
For an informal discussion of this property, see Appendix~\ref{app:lemma}.

Now, let $\mathcal{T}$ be a realization of the \ac{NTF} such that
\begin{multline}
  \left(\hspace{-0.5ex}
    \begin{array}{c|c}
      \mat A & \mat B\\[-0.5ex]
      \hlx{hv}
      \mat C & \mat D
    \end{array}
    \hspace{-0.5ex}\right)(z)=\\
  \left(\hspace{-0.5ex}
    \begin{array}{ccccc|c}
      0 & 1 & 0 & \cdots & 0 & 0 \\
      0 & 0 & 1 & \cdots & 0 & 0 \\
      \vdots & \vdots & \vdots & \ddots & \vdots & \vdots \\
      0 & 0 & 0 & \cdots & 1 & 0 \\
      0 & 0 & 0 & \cdots & 0 & 1 \\[-0.5ex]
      \hlx{hv}
      a_P & a_{P-1} & a_{P-2} & \cdots & a_1 & a_0
    \end{array}
    \hspace{-0.5ex}\right)(z) .
\end{multline}
This is a canonical realization, where $\mat A$ is responsible of making each
state variable a delayed version of the preceding one, so that the state
variables end up being a memory of the last $P$ input samples
\cite{Antoniou:DFAD-2000}. Evidently, this realization is minimal (thus
controllable) and only $\mat C$ depends on the filter coefficients that are
object of optimization. Hence, the left hand side of
inequality~\eqref{eq:matrix-inequality} is affine in the coefficients object of
minimization. Furthermore, it is \emph{affine} in the entries of $\mat
P$. Hence, collecting in a vector $\vec \xi$ all the filter coefficients $a_1,
\dots, a_P$ and all the independent entries in $\mat P$ to get an $L$ entry
vector $(\xi_1, \dots \xi_L)\transposed$, it must be
that~\eqref{eq:matrix-inequality} can be expressed as $\mat M(\vec \xi) \leq 0$
with
\begin{equation}
  \mat M(\vec \xi) = \mat M_0 + \sum_{i=1}^L \mat M_i\; \xi_i 
\end{equation}
where $\mat M_0, \dots \mat M_L$ are symmetric matrices. Regardless of the
entries of the $\mat M_i$ matrices (that are unimportant here), this shows
that~\eqref{eq:matrix-inequality} is a \ac{LMI}. Such property is quite
important, as it states that~\eqref{eq:matrix-inequality} is a convex
constraint.

\subsection{Summary of the optimization problem}
It is now possible to summarize the optimization problem.  To find $P$ filter
coefficients $a_1, \dots, a_P$, one needs to build a problem with $L$ variables
$\xi_1, \dots, \xi_L$. The first $P$ of them are the filter coefficients
themselves, while the last $L-P$ are entries of matrix $\mat P$, functional to
the solution of the problem, but uninteresting and due to be eventually
discarded.

The problem consists in the minimization of the convex quadratic form in
Eqn.~\eqref{eq:objective}, which is also a convex quadratic form in $\vec
\xi$. One has the constraint $\mat P>0$, which is an \ac{LMI} in $\vec \xi$ and
the constraint in Eqn.~\eqref{eq:matrix-inequality}, which has just been shown
to be an \ac{LMI} in $\vec \xi$. Altogether, one has a problem that can be
tackled by \ac{SDP}. In recent times, \emph{interior point methods}
\cite{Boyd:CO-2009,DeKlerk:ASP-2002} allow problems of this sort to be solved
in polynomial time with respect to the problem size. On commodity hardware,
problems with a thousand of variables or more can be solved in a few
seconds. In our case the number of variables is dominated by the number of
independent entries in $\mat P$, which is $P\times P$. Since $\mat P$ is
symmetric, this means having $\nicefrac{P}{2}\cdot(P+1)$ independent entries
and $\nicefrac{P}{2}\cdot(P+2)$ variables overall. Consequently, one can easily
go up to filter orders of $50$ or more. In practice, as
Section~\ref{sec:examples} shows, there is hardly a need to reach such high
orders. That section provides practical examples obtained with an \ac{SDP} code
distributed under a free, open source licence, proving that tackling our
optimization problem is not just possible, but also quite affordable.

\subsection{Positive side effects of the proposed optimization}
After having illustrated how the proposed \ac{NTF} optimization deals with
issue a) in Sec.~\ref{sec:issues}, it is worth considering also the other
items. Points b) and c) are automatically eliminated thanks to the different
design strategy. Particularly, the proposed methodology is completely agnostic
of the \ac{OSR}.
With respect to point d), our strategy tends to provide \acp{NTF} that are
\emph{only as steep as needed} and typically just as
steep as the reconstruction filter is.  Eventually, with
respect to point e), our strategy can deal with any kind of modulator, even the
most unusual one, with no need for frequency transformations. For instance, a
multi-band modulator will obviously have a multi-band output filter. Feeding
such filter into the our design procedure, automatically leads to the required
\ac{NTF}.

\section{Design examples and comparison to conventional design flows}
\label{sec:examples}
The design examples proposed in this Section have been tested by coding our
strategy in the Python programming language taking advantage of the Numpy and
Scipy packages \cite{Oliphant:CS+E-9-3}.  Python is a modern general purpose
programming language renown for its conciseness and extensibility. Numpy and
Scipy expand it into a powerful, matrix-oriented numerical computing
environment that can be freely deployed on all major computing platforms.
\ac{SDP} has been addressed using a further Python module, the CVXOPT free
software package for convex optimization
\cite{Andersen:OML-3-2011}. Specifically, CVXOPT has been used as a backend
solver, while the CVXPY package \cite{DeRubira:CVXPY-2012} has been employed as
a modeling framework to express the optimization problem in a more natural form
under the rules of \ac{DCP} \cite{Grant:GOTI-7-2006}. Comparison to
conventional ΔΣ modulator design flows has been practiced by the DELSIG toolbox
\cite{Schreier:DELSIG} and the code provided with
\cite{Nagahara:TSP-60-6}. DELSIG has also been used for the time domain
simulation of the modulators. A sample of our code is available. Please refer
to Appendix~\ref{app:code} for information on how to obtain it.

\subsection{\ac{LP} modulator with first order output filter}
\label{sec:lp1}
\begin{figure}[t]
  \begin{center}
      \includegraphics[scale=0.65]{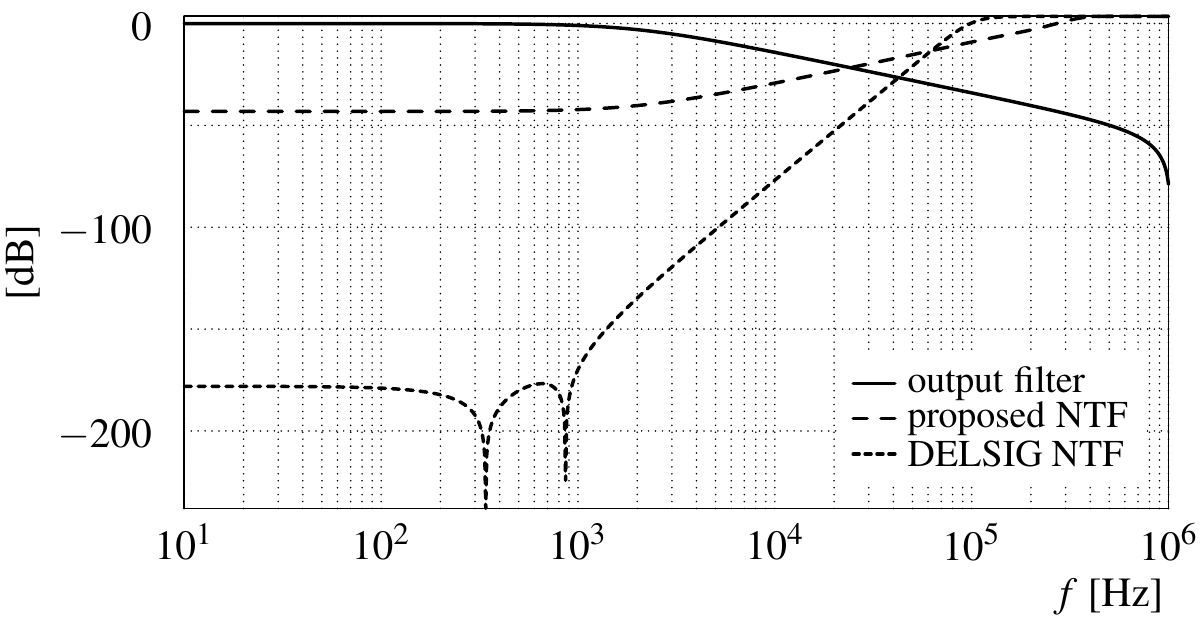}%
  \end{center}%
    \caption{Output filter, proposed \ac{NTF} and \ac{NTF} obtained by a
      conventional design flow, for the test case in Section~\ref{sec:lp1}.}
  \label{fig:lp1-filters}
\end{figure}

The first design case regards a binary \ac{LP} modulator, targeting signal
synthesis and coupled with a 1\Us{st} order reconstruction filter. The signal
band extends from dc to \unit[1]{kHz} and the \ac{OSR} is $1024$ (i.e., the
modulator operates at approximately \unit[2]{MHz}). To avoid spurious
attenuation at the filter output, the reconstruction filter is designed with
its cut-off frequency set at \unit[2]{kHz}. The Lee coefficient $\gamma$ is set
at $1.5$. Fig.~\ref{fig:lp1-filters} shows the output filter profile and the
\acp{NTF} obtained with the \texttt{synthesizeNTF} function in DELSIG (for a
4\Us{th} order modulator with optimized zeros) and our approach (for a
12\Us{th} order \ac{FIR} \ac{NTF}).\footnote{%
  We tried to compare also to the method in \cite{Nagahara:TSP-60-6}, but the
  provided code seems to run into numerical stability problems for this very
  large \ac{OSR}.}

\begin{figure}[t]
  \begin{center}
    \subfloat[\label{sfig:lp1-conv-shapes}]{%
      \includegraphics[scale=0.65]{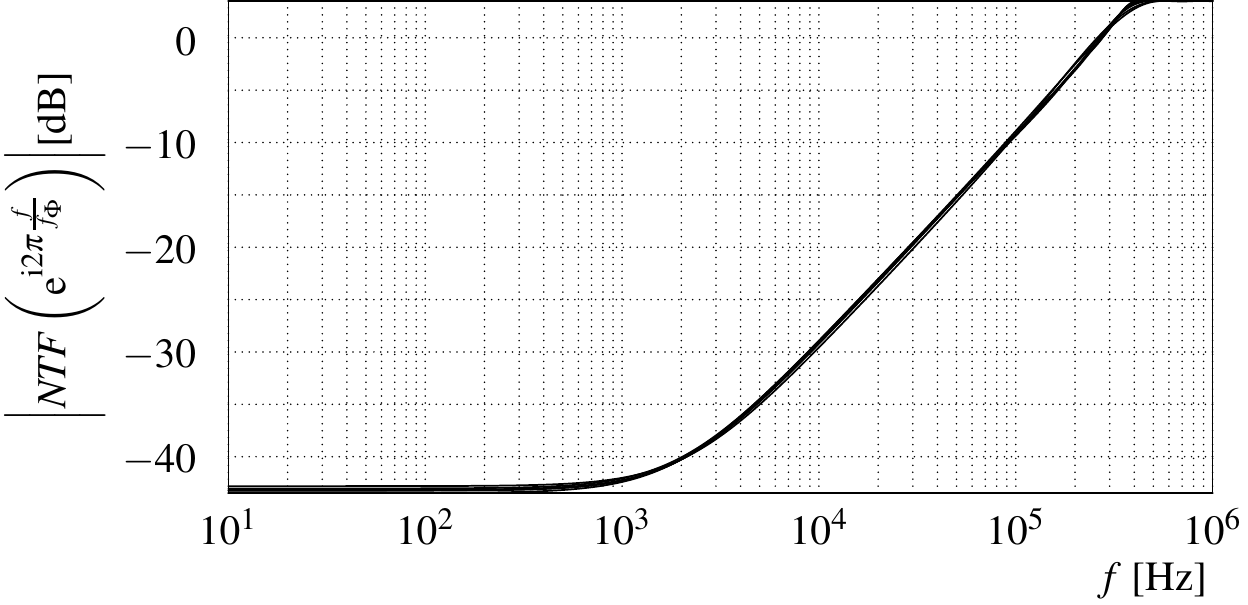}}\\
    \subfloat[\label{sfig:lp1-conv-perf}]{%
      \includegraphics[scale=0.65]{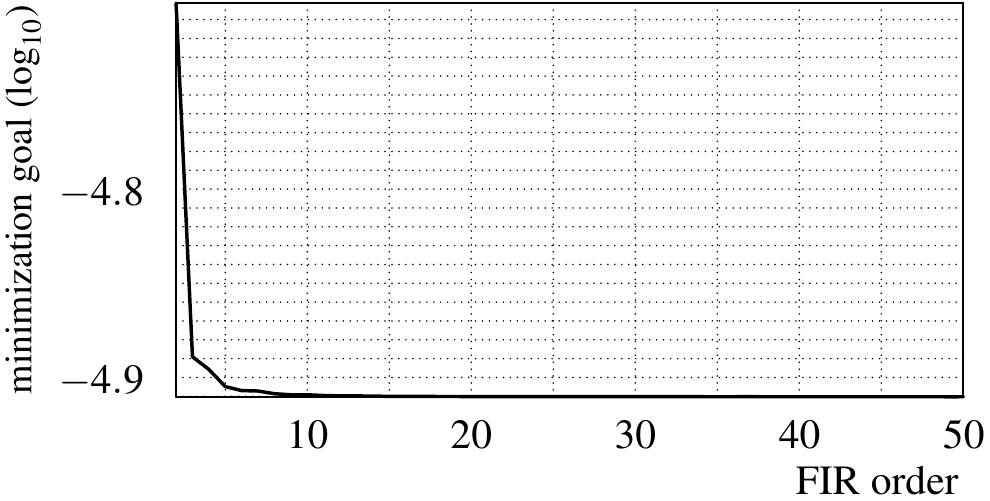}}    
  \end{center}
  \caption{Convergence of the \ac{NTF} to its optimal shape for increasing
    filter orders $5, 6, 9, 13, 18, 25$ \protect\subref{sfig:lp1-conv-shapes}
    and improvement in minimization goal as the \ac{FIR} order is increased
    \protect\subref{sfig:lp1-conv-perf}. In
    \protect\subref{sfig:lp1-conv-shapes}, the six curves superimpose almost
    perfectly. Data for the test case in Section~\ref{sec:lp1}.}
  \label{fig:lp1-convergence}
\end{figure}

The choice of a 12\Us{th} order for the proposed strategy is not casual. As a
matter of fact, in the proposed approach the performance improves (i.e.,
$\sigma_H$ reduces) as the order is increased. However, the improvement is
initially very rapid, then it slows down. This is well evident in figure
\ref{fig:lp1-convergence}, which shows the convergence to the optimal \ac{NTF}
shape. For orders higher than 6, the \ac{NTF} shape is almost
invariant. Clearly, it is convenient to stop increasing the order as soon as
$\sigma_H$ levels off, which happens slightly above 10. Interestingly, a
similar convergence is not experienced in other design strategies, which keep
delivering different (and improved, according to their merit factors) \acp{NTF}
as the order is increased, so that the limit is the loss of robust behavior or
the loss of numerical accuracy in the optimizer. Incidentally, this is the
reason why we can compare modulators having different orders. Indeed, we
compare the best modulator designed with the proposed strategy to modulators
designed with other strategies at a reasonable trade-off between quality and
robustness.

Interestingly, the \ac{NTF} obtained by our design flow turns out to be by far
less aggressive than the conventional one, exhibiting a much lower attenuation
in the signal band. Nonetheless, its performance is better. An estimation of
the output SNR, for an input sinusoid with amplitude $A=0.4$ (normalized to the
quantization levels set at $\pm 1$) can be obtained as
\begin{equation}
  \mathit{SNR}_{\text{expected}}=\frac{A^2}{2\sigma^2_H}
\end{equation}
and gives \unit[42.9]{dB} for the proposed approach and \unit[38.4]{dB} for the
conventional (\texttt{synthesizeNTF}) \ac{NTF} with a difference over
\unit[4.5]{dB}. Computing the SNR by time domain simulation (that let one use
the actual nonlinear modulator model) returns \unit[42.4]{dB} for the proposed
technique and \unit[40.3]{dB} for the conventional one, so that the advantage
reduces to about \unit[2]{dB}, still being well perceivable. The SNR numbers
have been obtained by replicating in software the architecture in
Fig.~\ref{fig:filtered-approx}. Excited with the modulator input alone as
$w(n)$, it lets one measure the signal power at $e(n)$, while excited with both
the modulator input at $w(n)$ and output at $x(n)$ it lets one measure the
noise power at $e(n)$.

\begin{figure}[t]
  \begin{center}
    \includegraphics[scale=0.65]{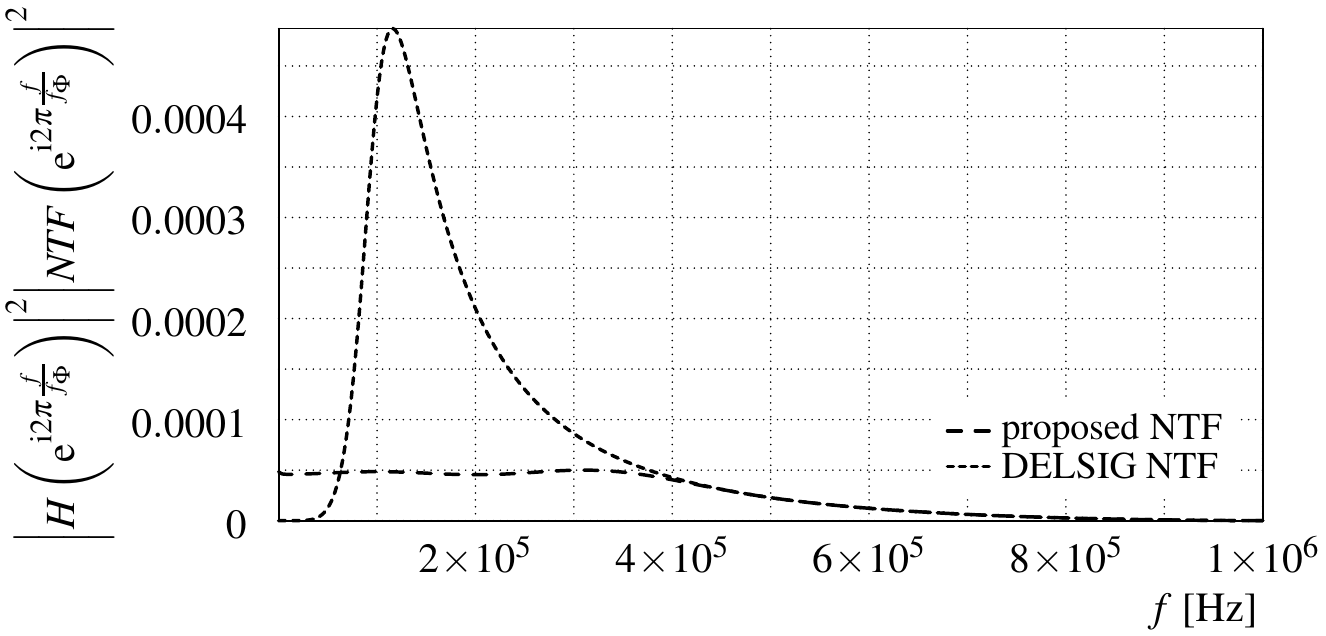}
  \end{center}%
    \caption{Integrand appearing in the definition of $\sigma^2_H$ for the
      proposed \ac{NTF} and the \acp{NTF} obtained by a conventional design
      flow. Data for the test case in Section~\ref{sec:lp1}.}
  \label{fig:lp1-ortho}
\end{figure}

A justification for the apparent paradox of having a better behavior with a
less aggressive \ac{NTF} comes from Fig.~\ref{fig:lp1-ortho}, which shows
$\abs{H\left(\ee^{\ii\omega}\right)}^2\abs{\NTF\left(\ee^{\ii\omega}\right)}^2$
in the three cases. Here, thanks to the linear scale, it is well evident that
the advantage of the conventional \acp{NTF} within the signal band is more than
compensated by the advantage of the proposed \ac{NTF} out of it.

It is even more interesting to note that a ΔΣ modulator based on the proposed
\ac{NTF} can behave much more robustly than a conventional one. Even with large
input signals, it can operate correctly, without overloading its
quantizer. Conversely, the 4\Us{th} order modulator obtained by the
\texttt{synthesizeNTF} design flow is much more fragile due to its
steepness. For instance, at a signal amplitude $A=0.7$ it already breaks,
unless the Lee coefficient $\gamma$ is lowered to $1.4$. Conversely, the
proposed \ac{NTF} makes the modulator work correctly up to $A=1$. Furthermore,
for $A$ values in little excess of $1$ where \emph{by definition} the modulator
is not meant to operate, one initially sees a graceful degradation of
performance, rather than a full breakage. For instance at $A=1.1$ one sees the
SNR reducing to \unit[30]{dB}.  The other way round, this increased robustness
can be used to bring the Lee coefficients to higher values without breaking the
modulator operation, cashing a further little advantage in terms of SNR. For
instance, for the output filter under exam, the proposed design technique lets
a modulator be designed with $\gamma=2$ gaining $1$ further \unit{dB} in
SNR. As a matter of fact, we have verified that even $\gamma=4$ is tolerated,
although without any SNR advantage.

Intuitively, the increased robustness is a consequence of the fact that the
proposed technique makes the \ac{NTF} no steeper than it is really needed,
matching the steepness of the output filter. Incidentally, this explains the
convergence in \ref{fig:lp1-convergence}. Once the required steepness is
reached, there is no need to rise the order any further.

\begin{figure}[t]
  \begin{center}
      \subfloat[\label{sfig:lp1-pz-opti}]{%
        \includegraphics[scale=0.65]{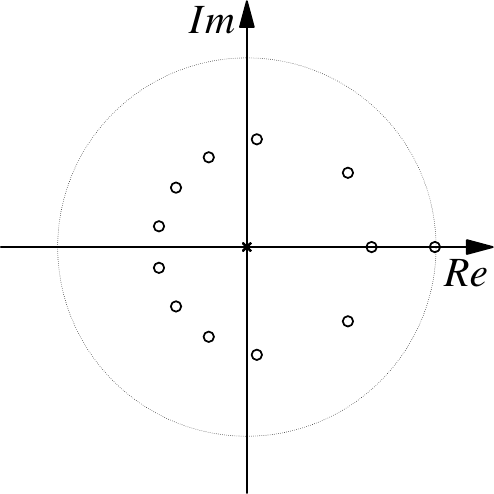}}\quad
      \subfloat[\label{sfig:lp1-pz-delsig}]{%
        \includegraphics[scale=0.65]{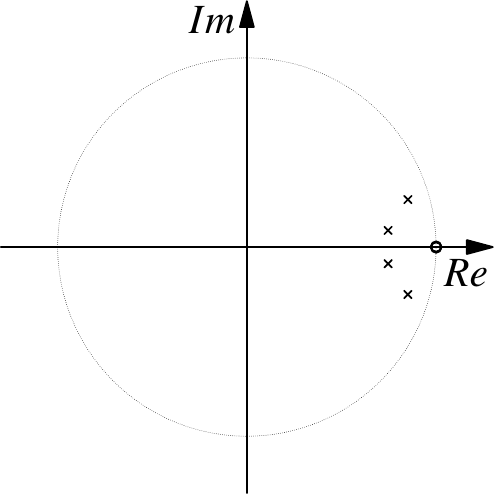}}%
  \end{center}%
    \caption{Comparison of pole-zero placement for the proposed design
      strategy~\protect\subref{sfig:lp1-pz-opti} and a conventional ones,
      namely \texttt{synthesizeNTF} from
      DELSIG~\protect\subref{sfig:lp1-pz-delsig}. Data from the
      test case in Section~\ref{sec:lp1}.}
  \label{fig:lp1-pz}
\end{figure}

Fig.~\ref{fig:lp1-pz} compares our pole/zero positioning to
conventional ones.

A final remark can be dedicated to the computational resources needed by the
optimization. For \ac{FIR} order $P=12$, a business laptop computer with a Core
2 Duo (Penryn, 2009) CPU and \unit[4]{GB} of RAM shared with the video card,
can perform the \ac{SDP} almost instantaneously.

\subsection{\ac{BP} modulator with output filter with steep
  features}\label{sec:bp1}
\begin{figure}[t]
  \begin{center}
      \includegraphics[scale=0.65]{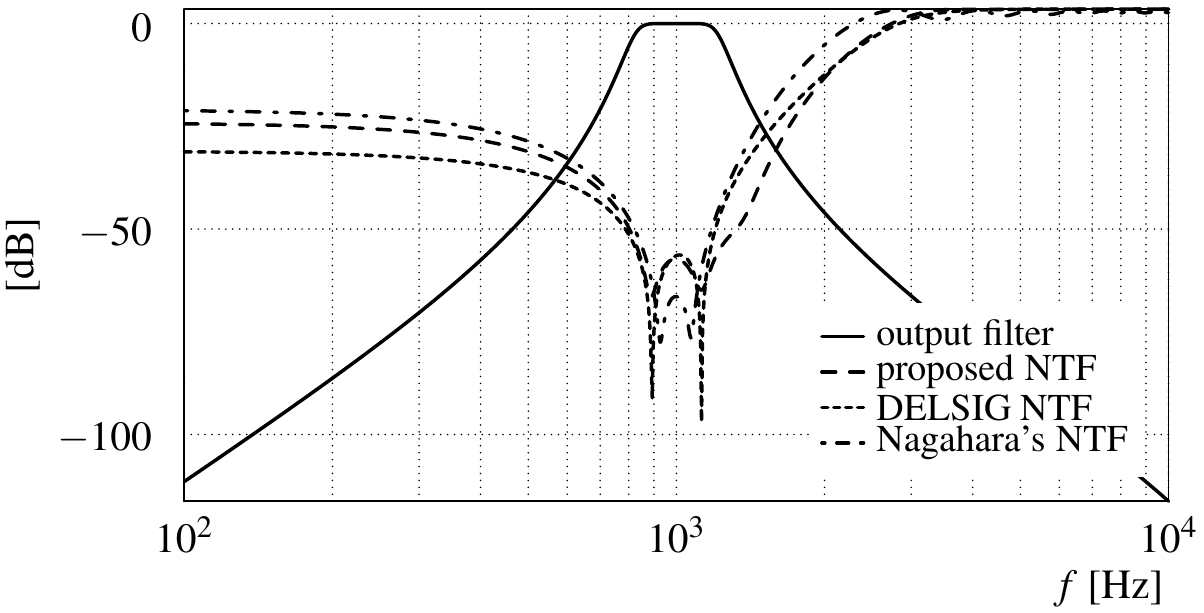}
  \end{center}
    \caption{Output filter, proposed \ac{NTF} and \acp{NTF} obtained by two
      conventional design flow, for the test case in Section~\ref{sec:bp1}.}
  \label{fig:bp1-filters}
\end{figure}

The second test case regards again a binary modulator, this time for \ac{BP}
signals. The signal band is centered at $\unit[1]{kHz}$ and extends for
$\unit[400]{Hz}$. The \ac{OSR} is set at 64. This time, the output filter is
much steeper than in the previous example, consisting in an 8\Us{th} order
Butterworth filter. The Lee coefficient $\gamma$ is set at
$1.5$. Fig.~\ref{fig:bp1-filters} shows the output filter profile and the
\acp{NTF} obtained with: the \texttt{synthesizeNTF} function in DELSIG (for a
4\Us{th} order modulator with optimized zeros); the method in
\cite{Nagahara:TSP-60-6} (for a 49\Us{th} order modulator); and our approach
(for a 49\Us{th} order \ac{FIR} \ac{NTF}).

\begin{figure}[t]
  \begin{center}
    \subfloat[\label{sfig:bp1-conv-shapes}]{%
      \includegraphics[scale=0.65]{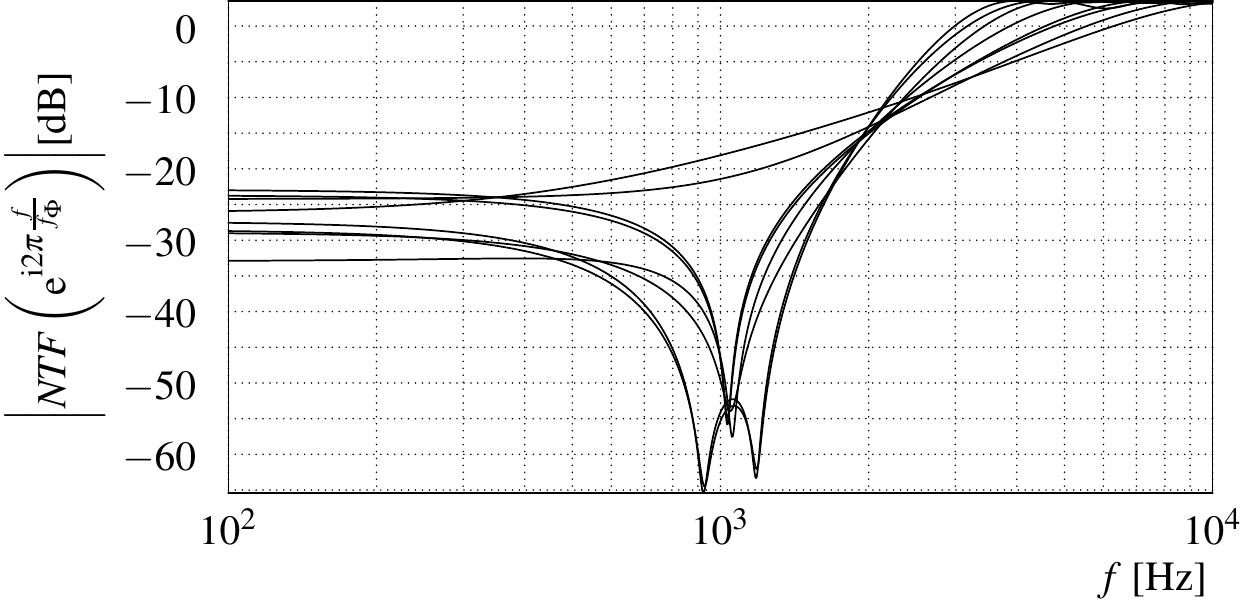}}\\
    \subfloat[\label{sfig:bp1-conv-perf}]{%
      \includegraphics[scale=0.65]{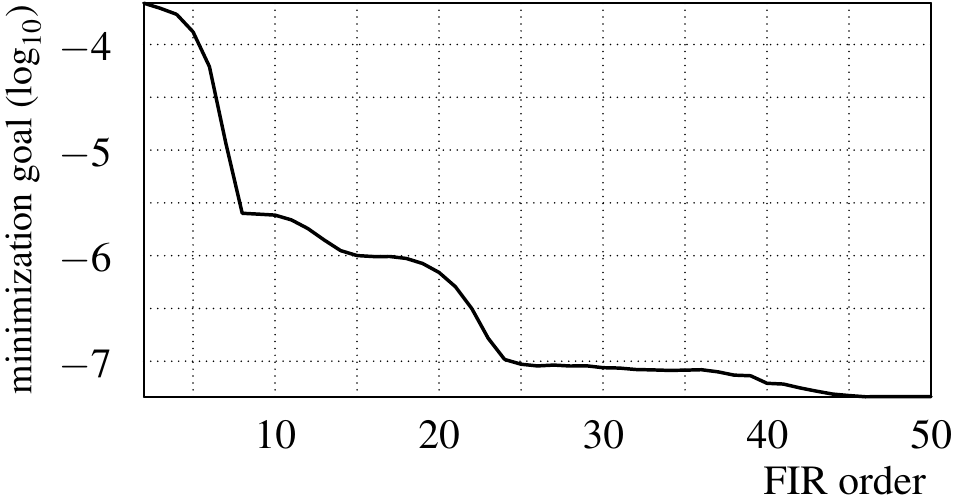}}
  \end{center}
  \caption{Convergence of the \ac{NTF} to its optimal shape for increasing
    filter orders $5, 8, 11, 15, 21, 28, 37, 49$
    \protect\subref{sfig:bp1-conv-shapes} and improvement in minimization goal
    as the \ac{FIR} order is increased \protect\subref{sfig:bp1-conv-perf}. In
    \protect\subref{sfig:bp1-conv-shapes}, the last 2 curves superimpose almost
    perfectly. Data for the test case in Section~\ref{sec:bp1}.}
  \label{fig:bp1-convergence}
\end{figure}

In this case, the higher roll-off of the output filter requires a higher
\ac{NTF} order for our methodology, as evident for the convergence analysis in
Fig.~\ref{fig:bp1-convergence}. From the second plot, a 32\Us{th} order
\ac{NTF} would already give good results. Note that the 49\Us{th} order
\ac{FIR} takes a couple of minutes to compute via \ac{SDP} on the same laptop
used for the previous test case.

\begin{figure}[t]
  \begin{center}
    \includegraphics[scale=0.65]{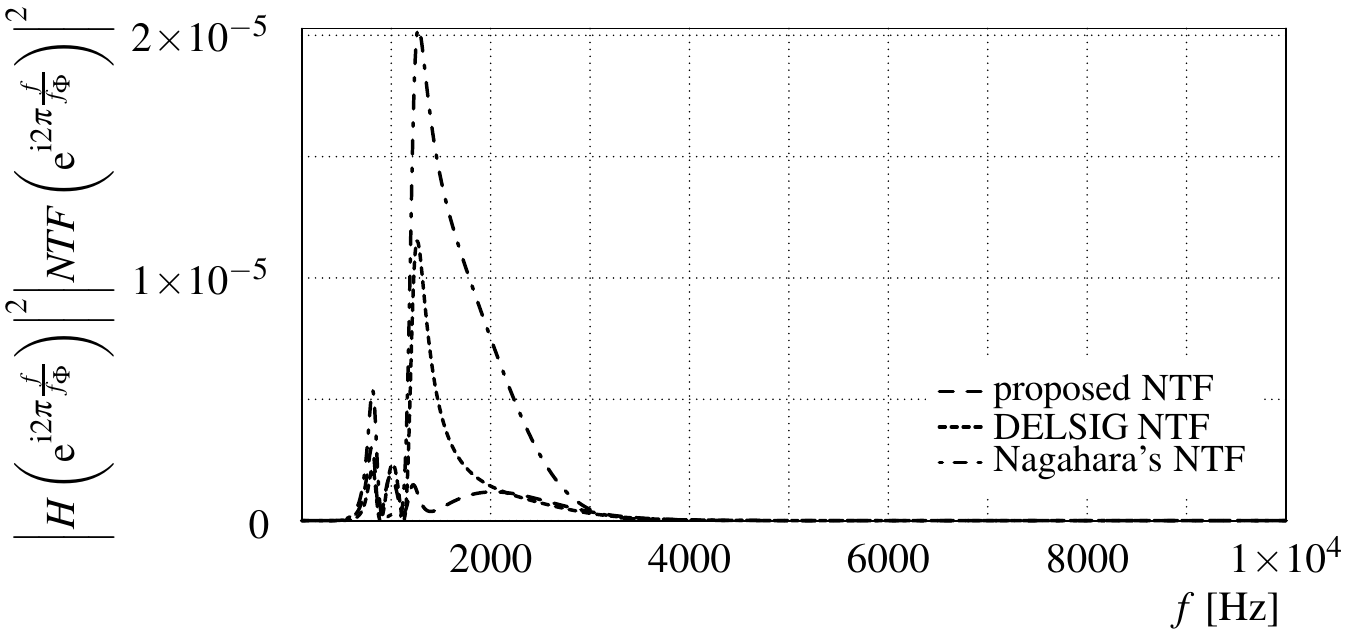}%
  \end{center}%
  \caption{Integrand appearing in the definition of $\sigma^2_H$ for the
    proposed \ac{NTF} and the \acp{NTF} obtained by two conventional design
    flows. Data for the test case in Section~\ref{sec:bp1}.}
  \label{fig:bp1-ortho}
\end{figure}

As for the previous test case, it is interesting to observe the integrand
appearing in the definition of $\sigma^2_H$. This is plotted in
Fig.~\ref{fig:bp1-ortho}.

In this case, for a sinusoidal input with $A=0.75$ we get an SNR (from time
domain simulation) after the output filter of \unit[69.2]{dB} for our design
approach, \unit[67.0]{dB} for the \texttt{synthesizeNTF} design flow, and
\unit[68.1]{dB} for the method in \cite{Nagahara:TSP-60-6}. Note that trying to
pick a 6\Us{th} order \ac{NTF} with the \texttt{synthesizeNTF} design flow
would lead to a misbehaving modulator, while our approach enables increasing
the \ac{FIR} order even above $49$.

\begin{figure}[t]
  \begin{center}
    \subfloat[\label{sfig:bp1-pz-opti}]{%
      \includegraphics[scale=0.65]{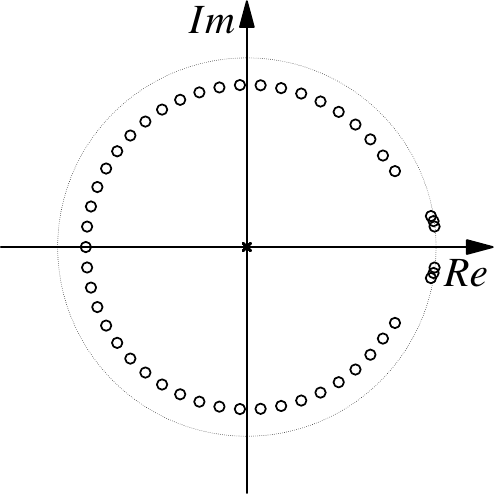}}\quad
    \subfloat[\label{sfig:bp1-pz-delsig}]{%
      \includegraphics[scale=0.65]{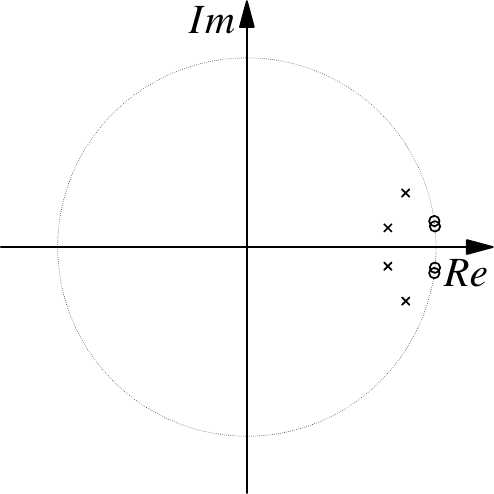}}\\[-1cm]
    \subfloat[\label{sfig:bp1-pz-naga}]{%
      \includegraphics[scale=0.65]{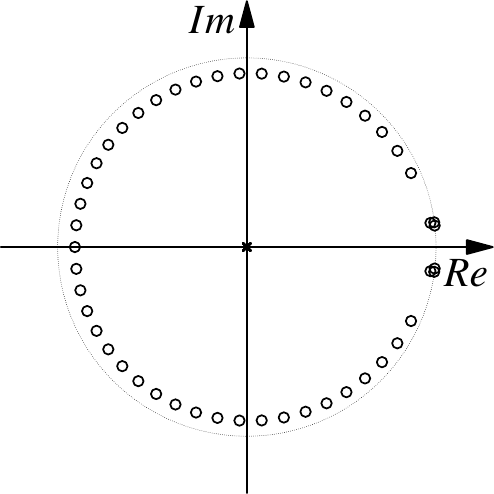}}
  \end{center}
  \caption{Comparison of pole-zero placement for the proposed design
    strategy~\protect\subref{sfig:bp1-pz-opti} and two conventional ones,
    namely \texttt{synthesizeNTF} from
    DELSIG~\protect\subref{sfig:bp1-pz-delsig} and Nagahara's strategy in
    \cite{Nagahara:TSP-60-6}~\protect\subref{sfig:bp1-pz-naga}. Data from the
    test case in Section~\ref{sec:bp1}.}
  \label{fig:bp1-pz}
\end{figure}

Fig.~\ref{fig:bp1-pz} compares our pole/zero positioning to the conventional
ones for this test case.

This test case shows that the advantage of the proposed approach may fade a
little when the output filter has steep cutoff characteristics close to an
ideal on-off behavior. This is quite reasonable, since an on-off filter
behavior is exactly the premise on which conventional design flows are
funded. Yet, even in this case some advantages remain evident, including those
in terms of SNR.

\subsection{Multi-band modulator}%
\label{sec:bpm}%
The last case that we consider is that of a multi-band modulator. This is
intractable in many conventional design flows \cite{Schreier:DELSIG}, although
it can be managed by the recent methodology in
\cite{Nagahara:TSP-60-6}.\footnote{Note that the method in
  \cite{Nagahara:TSP-60-6} targets a slightly different goal for the single and
  the multi-band case. Furthermore, it requires a new matrix inequality for
  each band so that it can become increasingly demanding in terms of
  computational power when their number is increased. Finally, the sample code
  delivered with the paper cannot deal with cases where the signal bands have
  different widths, although it can be easily extended for it.} %
Assume that the input signal has two bands, one centered at \unit[1]{kHz} and
\unit[400]{Hz} wide and the other centered at \unit[10]{kHz} and \unit[4]{kHz}
wide. Let the \ac{OSR} be 64 (i.e., $f_\Phi=\unit[2\cdot 64\cdot
(4000+400)]{Hz}$). Consider the case of a 2-band 8\Us{th} order Butterworth
filter at the output. As usual, for the modulator design consider the binary
case, with $\gamma=1.5$. Fig.~\ref{fig:bp1-filters} shows the output filter
profile and the \acp{NTF} obtained with in our approach (for a 50\Us{th} order
\ac{FIR} \ac{NTF}). Obviously, it is not possible to design an \ac{NTF} for
this case using \texttt{synthesizeNTF}, so we provide comparison only to the
the method in \cite{Nagahara:TSP-60-6} for the same \ac{FIR} order.

\begin{figure}[t]
  \begin{center}
    \includegraphics[scale=0.65]{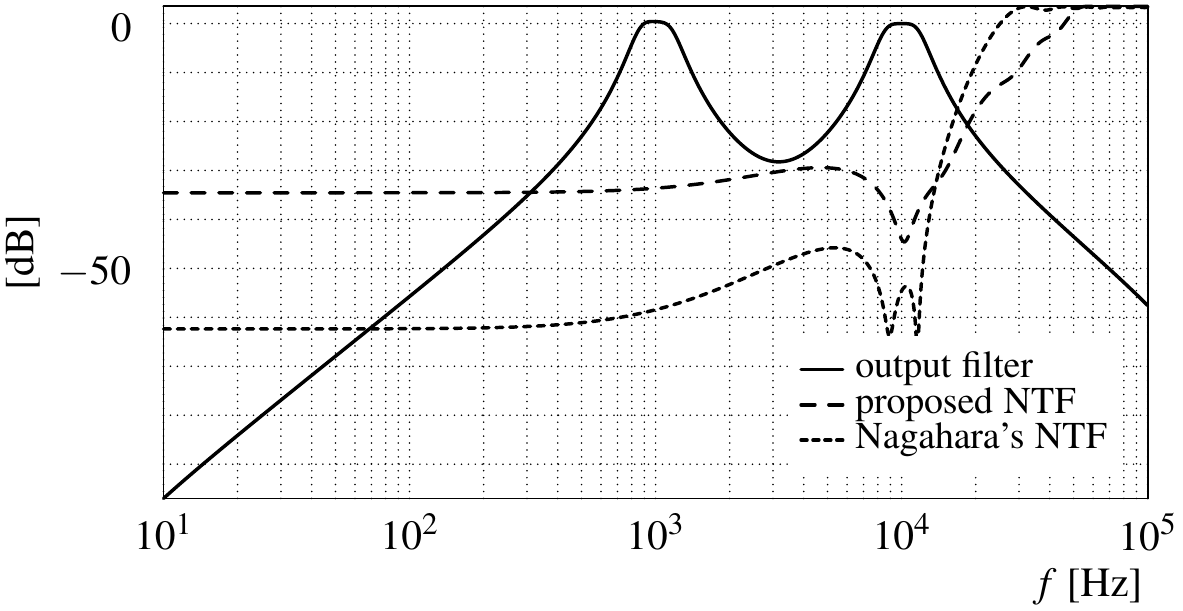}
  \end{center}
    \caption{Output filter, proposed \ac{NTF}, and conventional \ac{NTF}
      synthesized by the method in \cite{Nagahara:TSP-60-6} for the test case
      in Section~\ref{sec:bpm}.}%
  \label{fig:bpm-filters}
\end{figure}

\begin{figure}[t]
  \begin{center}
    \subfloat[\label{sfig:bpm-conv-shapes}]{%
      \includegraphics[scale=0.65]{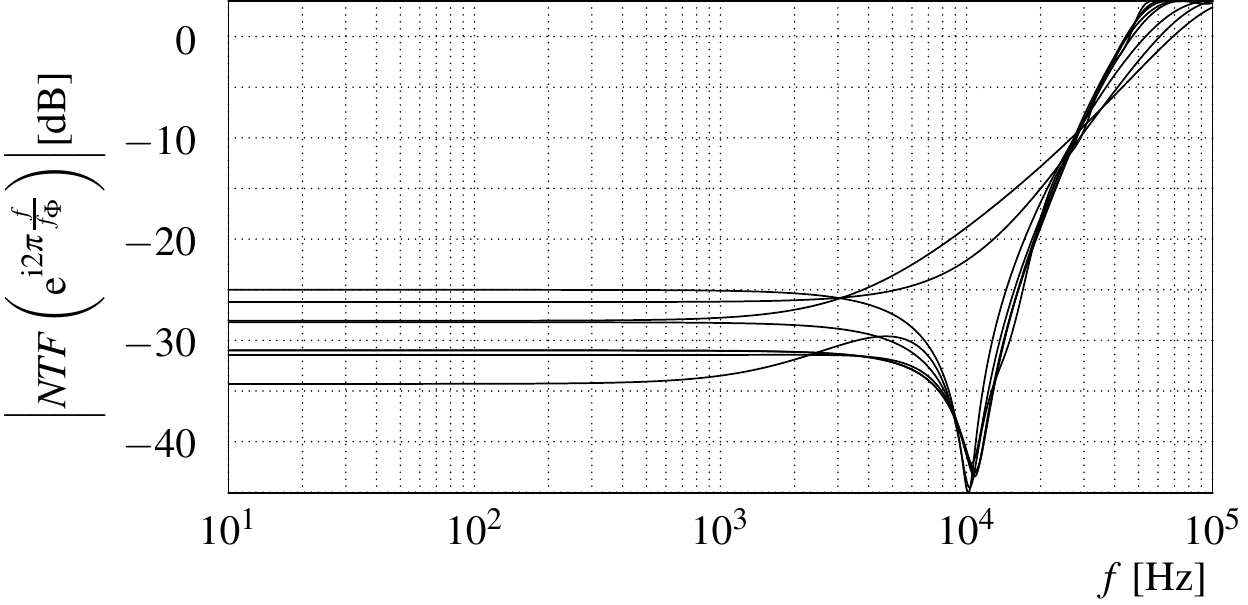}}\\
    \subfloat[\label{sfig:bpm-conv-perf}]{%
      \includegraphics[scale=0.65]{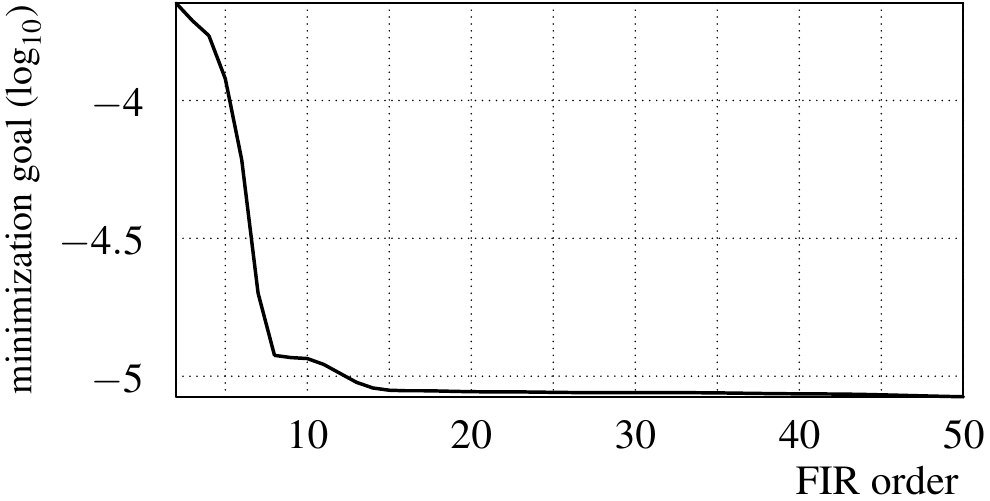}}
  \end{center}
  \caption{Convergence of the \ac{NTF} to its optimal shape for increasing
    filter orders $5, 8, 11, 15, 21, 28, 37, 49$
    \protect\subref{sfig:bpm-conv-shapes} and improvement in minimization goal
    as the \ac{FIR} order is increased \protect\subref{sfig:bp1-conv-perf}. In
    \protect\subref{sfig:bpm-conv-shapes}, the last 3 curves superimpose very
    well. Data for the test case in Section~\ref{sec:bpm}.}
  \label{fig:bpm-convergence}
\end{figure}

As for the previous test cases, Fig.~\ref{fig:bpm-convergence} shows the
\ac{NTF} convergence to the optimal shape as the \ac{FIR} order $P$ is
increased. Evidently, \ac{FIR} orders of 16 would already be enough to achieve
a good SNR.

In this case, simulating the modulator for an input signal given by the
superposition of two tones at 1 and 10 \unit{kHz}, with amplitude $A=0.45$ for
both of them, gives an SNR of over \unit[46]{dB} at the filter output. The
modulator based on the method in \cite{Nagahara:TSP-60-6} is unstable at this
large input. At $A=0.40$ it operates correctly, though, and it can be taken as
a reference. In this condition, our method delivers a \unit[48.2]{dB} SNR, vs a
\unit[42.3]{dB} SNR for the reference algorithm, namely we have an almost
\unit[6]{dB} advantage. This large advantage should be no surprise, since we
explicitly optimize for the SNR on the filtered output, while
\cite{Nagahara:TSP-60-6} uses a different merit factor.

\begin{figure}[t]
  \begin{center}
    \includegraphics[scale=0.65]{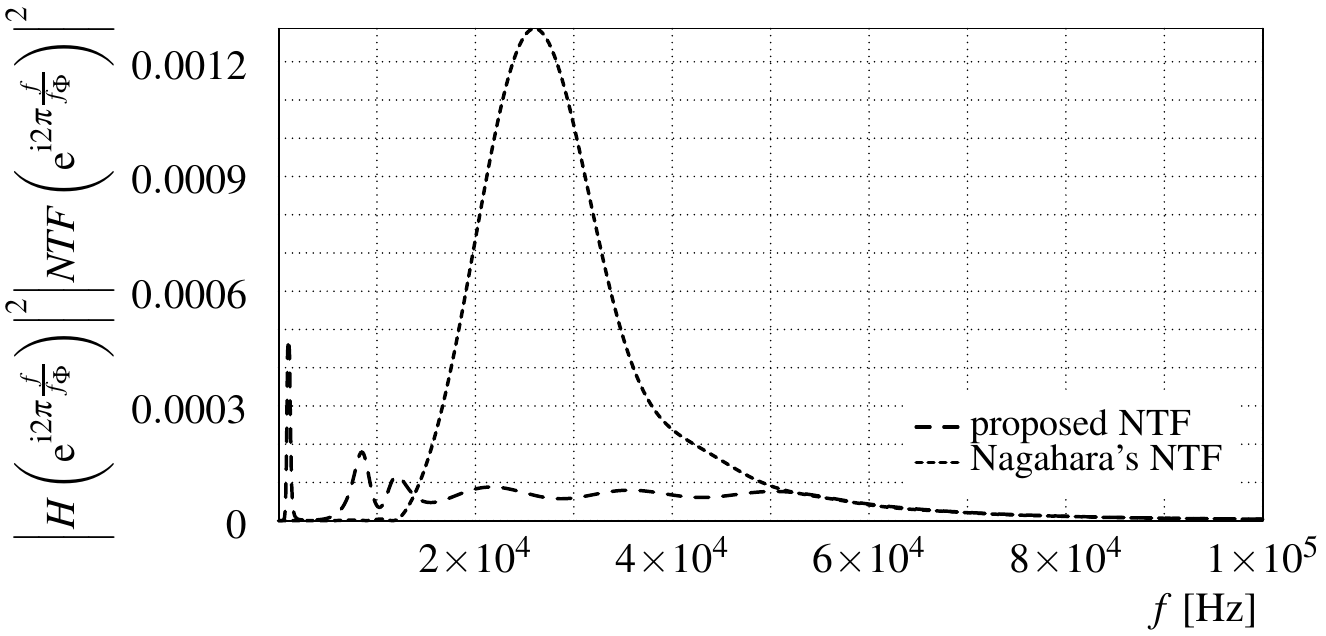}%
  \end{center}%
  \caption{Integrand appearing in the definition of $\sigma^2_H$ for the
    proposed \ac{NTF} and a conventional \ac{NTF} designed following the
    method in \cite{Nagahara:TSP-60-6}. Data for the test case in
    Section~\ref{sec:bpm}.}%
  \label{fig:bpm-ortho}
\end{figure}

What is interesting about this multi-band case is that it shows a rather
counter-intuitive behaviour. Looking at the input signal structure, one would
probably think that the modulator should put its quantization noise in 3
regions: at low frequencies, before the first signal band, at intermediate
frequencies, between the signal bands and at high frequencies, above the second
signal band. Furthermore, one would think that the modulator should have zones
of very high attenuation for the noise in the two signal bands.  Conversely,
our design approach shows that it is more convenient to use all the available
degrees of freedom on the \ac{NTF} to optimize the noise shaping at the high
frequencies, completely ignoring the two lower bands that are anyway extremely
thin and thus incapable to contain much noise. Additionally, it shows that it
would be a waste to strive to remove too much noise from the first signal band,
that is anyway very thin and thus incapable to contribute much to the overall
SNR.  This is very well evident from the graph in Fig.~\ref{fig:bpm-ortho}
which shows
$\abs{H\left(\ee^{\ii\omega}\right)}^2\abs{\NTF\left(\ee^{\ii\omega}\right)}^2$,
namely the integrand in the expression of $\sigma^2_H$, both for our \ac{NTF}
and the one obtained following \cite{Nagahara:TSP-60-6}. Thanks to the linear
scale, it is apparent that it is much more important to practice a good noise
allocation at the high frequencies above \unit[12]{kHz} than in the thin bands
between dc and \unit[800]{Hz} and between 1200 and
\unit[8000]{Hz}. Furthermore, it is interesting to look at the first peak in
the plot. This is due to the fact that the \ac{NTF} attenuates much less in the
first signal band than in the second. However, the linear scale makes this peak
appear as it really is: so thin that its \emph{mass} and thus its contribution
to the overall SNR is anyway very modest. Indeed, the \ac{NTF} based on
\cite{Nagahara:TSP-60-6}, that strives to remove a lot of noise also from the
first signal band, lacks this peak, but pays it with a much higher integrand
right above the second band.

\begin{figure}[t]
  \begin{center}
    \subfloat[\label{sfig:bpm-pz-opti}]{%
      \includegraphics[scale=0.65]{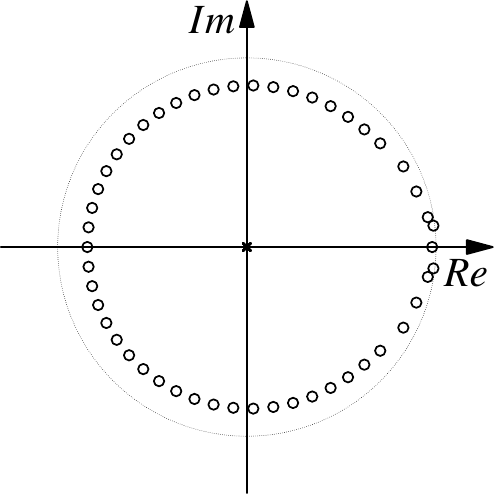}}\quad
    \subfloat[\label{sfig:bpm-pz-naga}]{%
      \includegraphics[scale=0.65]{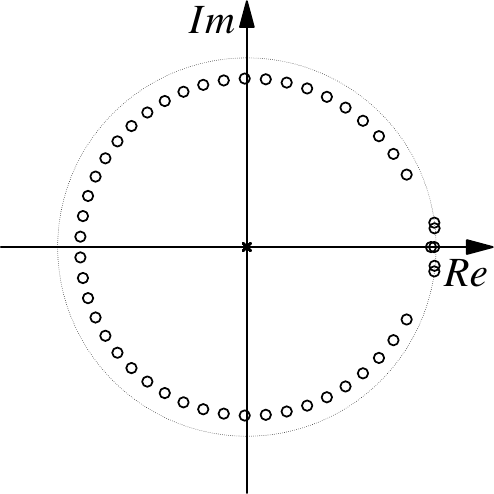}}
  \end{center}
  \caption{Pole-zero placement for the proposed design
    strategy~\protect\subref{sfig:bpm-pz-opti} and the conventional one in
    \cite{Nagahara:TSP-60-6}~\protect\subref{sfig:bpm-pz-naga} for the test
    case in Section~\ref{sec:bpm}.}
  \label{fig:bpm-pz}
\end{figure}

Finally, Fig.~\ref{fig:bpm-pz} shows the pole-zero location, which is somehow
similar to that in Fig.~\ref{fig:bp1-pz}, given that also in this case we end
up with a \ac{BS} \ac{NTF}.

\section{Conclusions}
\label{sec:conclusions}
In this paper we have proposed a new design flow for ΔΣ modulators. Contrarily
to conventional strategies, our methodology is aware of the output filter that
in most practical applications is placed at the modulator output. This is an
important difference, since common strategies assume that an ideal filter is
available, with a steep on-off behavior between the signal band and the noise
band. Consequently, they only strive at putting most of the noise out of the
signal band. Conversely, our strategy strives to shape the quantization noise
so that the overall noise after the output filter can be minimized.

In practice the two approaches may become similar when a steep filter is indeed
available. Yet, they significantly differ when one can only count on a filter
with a non-ideal roll-off. In this latter case, our strategy consistently
provides a better behavior. It is worth underlining that this situation emerges
in a large number of applications. Specifically, whenever one uses ΔΣ
modulation for signal synthesis, power conversion or actuation, the output
filter is analog and often lacks aggressive specifications. As a matter of
fact, when one talks about actuation (ac motor drives, audio amplification), it
frequently happens that the filter is realized taking advantage of the actuator
input impedance. In this case, the filter designer may get quite confined in
his choices. Extensive simulations (some of which presented in the paper) have
shown evident SNR improvements in cases where the modulator works in
conjunction with not-so-good output filters. However, even when output filters
with a good roll-off are available there might be minor SNR increases.

Another distinguished advantage of the proposed approach is that it does never
lead to a \ac{NTF} that is steeper than it is strictly needed. This helps
keeping the modulator stable and resilient to input conditions that could
otherwise undermine its behavior (e.g., too large input signals).

Finally, the proposed methodology lets one tackle design cases that are often
unmanageable by conventional flows. As an example, we have proposed the case of
a modulator for multi-band signals.

The approach that we have described is based on the \ac{KYP} lemma and obtains
the \ac{NTF} through an optimization process exploiting \ac{SDP}. The resulting
\acp{NTF} are in the \ac{FIR} class. Current day algorithms enable a very rapid
design process even for high-order modulators. Typical computation times may be
between a few seconds to a few minutes even on a standard laptop. In practice,
our simulations show that, in most cases, \ac{FIR} orders can be kept
relatively low.

As a final remark, note that being based on a formal process, our strategy can
return truly optimal \ac{NTF} shapes. Interestingly, there are cases when these
may at first appear counter-intuitive so that only a more thorough exam lets
one see a justification.
 
\appendix
\subsection{Discussion about the bounded real lemma}
\label{app:lemma}
Here, we provide an informal discussion of
Property~\ref{prop:bounded-real}. The interested reader is invited to refer to
\cite{Boyd:LMISCT-1994} for formal considerations.

The so called \ac{KYP} lemma actually refers to a loosely defined set of
theorems related to dissipation inequalities.  Restricting to discrete time
models, consider a system $\mathcal{T}$ with $n$ state variables $\vec x$, $m$
inputs $\vec u$ and $p$ outputs $\vec y$
\begin{equation}
  \mathcal{T}=
  \left(\hspace{-0.5ex}
    \begin{array}{c|c}
      \mat A & \mat B\\[-0.5ex]
      \hlx{hv}
      \mat C & \mat D
    \end{array}
    \hspace{-0.5ex}\right)(z)
  \label{eq:model-T}
\end{equation}

The system is said to be dissipative with respect to a real valued supply rate
$s(\vec x, \vec u)$ if there exists a continuous real valued function $V(\vec
x)$, named storage function, such that the dissipation inequality
\begin{equation}
  V(\vec x(n+1))-V(\vec x(n)) \le s(\vec x(n), \vec u(n))
\end{equation}
holds for all the admissible $\vec x$, $\vec u$ trajectories. Clearly, the
storage function can be seen as a generalization of the energy accumulated by
the system, while the supply rate can be seen as a generalization of the rate
at which energy is provided to or taken from it.

For practical systems, it is reasonable to assume that the rate function is a
quadratic form on $\vec x$ and $\vec u$, as in
\begin{equation}
  s(\vec x, \vec u) = 
  \begin{pmatrix}
    \vec x\\ \vec u
  \end{pmatrix}\transposed\! 
  \hat{\mat Q} 
  \begin{pmatrix}
    \vec x\\ \vec u
  \end{pmatrix}
  \label{eq:quadratic-supply}
\end{equation}
where $\hat{\mat Q}$ is a symmetric real matrix.

A first part of the \ac{KYP} lemma establishes that for a linear system with a
quadratic supply rate the dissipation inequality can be satisfied for some
continuous storage function $V(\vec x)$ if and only if it is satisfied for some
\emph{quadratic} storage function $V(\vec x) = \vec x\transposed \mat P \vec x$
where $\mat P$ is real symmetric. In this case, the dissipation inequality can
be expressed as
\begin{equation}
  (\mat A \vec x + \mat B \vec u)\transposed\,\mat P\,
  (\mat A \vec x + \mat B \vec u)  -
  \vec x\transposed\,\mat P\,\vec x \le s(\vec x, \vec u)
  \label{eq:dissipation-inequality}
\end{equation}
This can be equivalently recast as
\begin{multline}
  \begin{pmatrix}
    \vec x\\ \vec u\end{pmatrix}\transposed
  \begin{pmatrix}
    \mat A\transposed\, \mat P\, \mat A - \mat P & 
    \mat A\transposed\, \mat P\, \mat B\\
    \mat B\transposed\, \mat P\, \mat A &
    \mat B\transposed\, \mat P\, \mat B
  \end{pmatrix}\,
  \begin{pmatrix}
    \vec x\\ \vec u
  \end{pmatrix}
  \le\\ 
  \begin{pmatrix}
    \vec x\\ \vec u
  \end{pmatrix}\transposed 
  \hat{\mat Q} 
  \begin{pmatrix}
    \vec x\\ \vec u
  \end{pmatrix}
\end{multline}

Eventually, since the inequality should hold for all $\vec x$ and $\vec u$ one
has
\begin{equation}
  \begin{pmatrix}
    \mat A\transposed \mat P \mat A - \mat P & 
    \mat A\transposed \mat P \mat B\\
    \mat B\transposed \mat P \mat A &
    \mat B\transposed \mat P \mat B
  \end{pmatrix} - \hat{\mat Q} \le 0
  \label{eq:gmi}
\end{equation}
which should be intended as a generalized matrix inequality imposing negative
semi-definiteness. Furthermore, posing a positive definiteness restriction on
$V(\vec x)$ requires and is implied by $\mat P$ being positive semi-definite.

A second part in the lemma establishes that while originally conceived for real
state and input vector $\vec x$ and $\vec u$, it can be extended to complex
vectors $\vec X$ and $\vec U$. With this, Eqn.~\eqref{eq:quadratic-supply}
becomes a \emph{Hermitian} form. Interestingly, by taking $\vec X$ and $\vec U$
so that model~\eqref{eq:model-T} is valid in the $z$-domain and restricting to
$z \in \Nset{C}$ such that $\abs{z} = 1$ (i.e., to sinusoidal regime with
$\omega \in [-\pi, \pi]$ and $z=\ee^{\ii\omega}$), the
inequality~\eqref{eq:dissipation-inequality} compells the Hermitian form to be
positive.  Most notably, the \emph{frequency domain inequality} version of the
\ac{KYP} states that the inverse also holds. Namely, the positive
semi-definiteness of $s(\vec X, \vec U)$ with complex variables and $z =
\ee^{\ii\omega}$ requires the existence of a positive semi definite real
symmetric $\mat P$ satisfying~\eqref{eq:gmi}. Actually, for this part of the
lemma to hold an additional assumption of controllability on $\mathcal{T}$ is
required.

Let us now take $m$ and $p$ equal to $1$ and assume that
\begin{equation}
  s(\vec X, U) = \gamma\abs{U}^2 - \abs{\mat C \vec X + \mat D U}^2
  \label{eq:pre-hermitian}
\end{equation}
Restricting to the unit circle in the $z$ plane, the positive semi-definiteness
of such an $s(\cdot)$ means that at any frequency $\omega$ the power of $y
=\mat C \vec x + \mat D u$ is less than $\gamma^2$ times the power of $u$,
namely that the input-output gain of the system is less than $\gamma$. In other
words, if $T(z)$ is the transfer function between $u$ and $y$, one has
$\norm{T}_\infty \leq \gamma$.

The Hermitian form associated to Eqn.~\eqref{eq:pre-hermitian} needs
\begin{equation}
  \hat{\mat Q} = 
  \begin{pmatrix}
    -\mat C\transposed \mat C & -\mat C\transposed \mat D\\
    -\mat D\transposed \mat C & -\mat D\transposed \mat D +\gamma^2
  \end{pmatrix}
\end{equation}
Substituting it into~\eqref{eq:gmi} gives
\begin{equation}
  \begin{pmatrix}
    \mat A\transposed \mat P \mat A - \mat P + \mat C\transposed \mat C & 
    \mat A\transposed \mat P \mat B + \mat C\transposed \mat D \\
    \mat B\transposed \mat P \mat A + \mat D\transposed \mat C &
    \mat B\transposed \mat P \mat B - \gamma^2 + \mat D\transposed \mat D 
  \end{pmatrix} \le 0
  \label{eq:gmi2}
\end{equation}

Let us now look at the matrix
\begin{equation}
  \begin{pmatrix}
    \mat A\transposed\,\mat P\,\mat A - \mat P &
    \mat A\transposed\,\mat P\,\mat B &
    \mat C\transposed\\
    \mat B\transposed\,\mat P\,\mat A &
    \mat B\transposed\,\mat P\,\mat B - \gamma^2&
    \mat D\\
    \mat C & \mat D & -1
    \label{eq:complementable}
  \end{pmatrix} .
\end{equation}
Evidently, the matrix in~\eqref{eq:gmi2} is the \emph{Schur's complement} of
the bottom right sub-matrix $(-1)$ in
matrix~\eqref{eq:complementable}. Eventually, recall the Schur's complement
conditions for positive (negative) definiteness \cite{Zhang:SCA-2005}. These
basically state that a matrix $\mat M$ is positive (negative) semi-definite if
and only if the Schur's complement $\mat S$ of a sub-matrix $\mat S_M$ in $\mat
M$ is positive (negative) semi-definite and $\mat S_M$ is positive (negative)
definite. This implies that being $-1$ a negative scalar, the inequality
$\eqref{eq:gmi2}$ holds if and only if the matrix in \eqref{eq:complementable}
is negative semi-definite.

\subsection{Sample code for the proposed design method}
\label{app:code}
Sample code for the proposed ΔΣ Modulator design flow can be downloaded at the
following site \url{http://pydsm.googlecode.com}. The code is licensed
under a free, open-source license. Please follow the instructions in the
\texttt{README} to install the software and all the dependencies necessary for
its usage. The \texttt{README} also contains some information on how to
replicate the examples proposed in this paper. In case you find this software
useful, please propagate information about this paper which constitutes a
fundamental part of its documentation.


\ifCLASSOPTIONcaptionsoff
  \newpage
\fi



\bibliographystyle{IEEEtran}
\bibliography{macros,IEEEabrv,various,chaos,analog}
%
%
%
\begin{IEEEbiography}[{%
    \includegraphics[width=1in,height=1.25in,clip,keepaspectratio]{%
      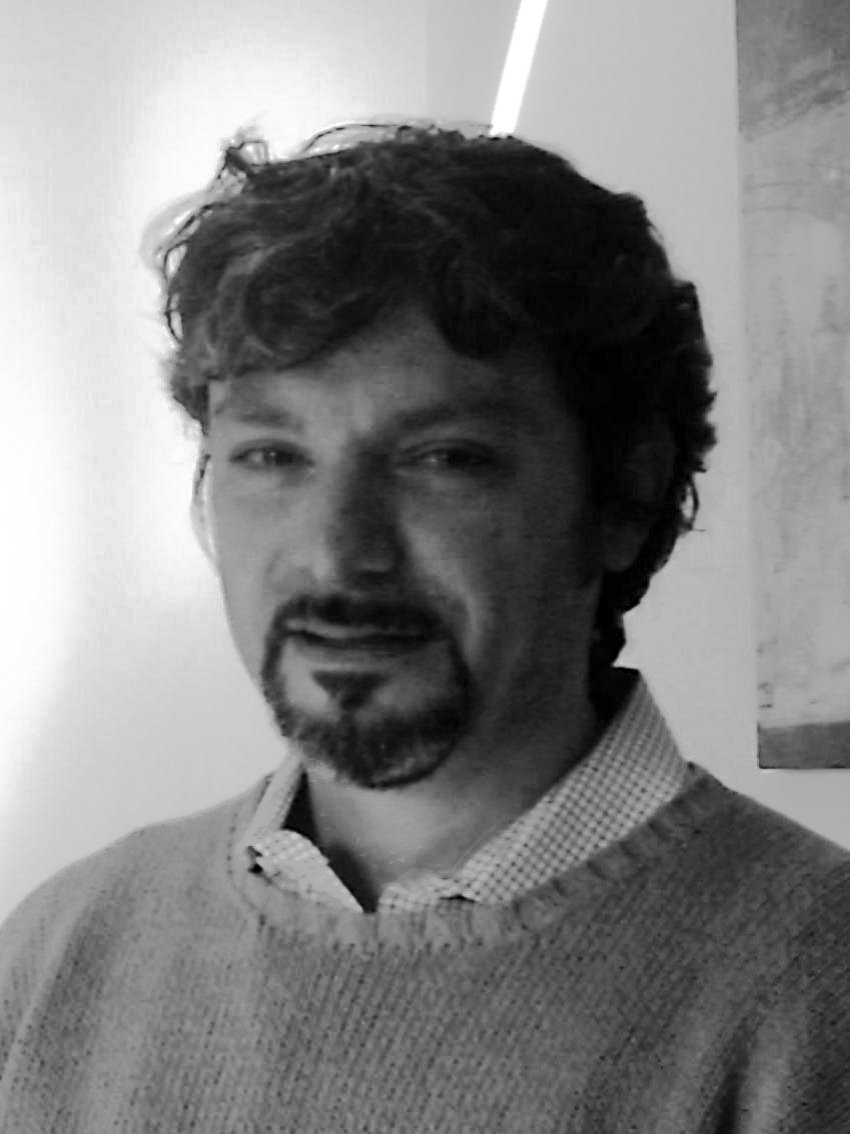}}]{%
    Sergio Callegari}%
  (M'00--SM'06) received a Dr.~Eng.\@ degree (with honors) in electronic
  engineering and a Ph.D.\@ degree in electronic engineering and computer
  science from the University of Bologna, Italy, in 1996 and 2000,
  respectively, working on the study of nonlinear circuits and chaotic
  systems. In 1996, he was a visiting student at King’s College London,
  UK. He is currently a researcher and assistant professor at the School
  of Engineering II, University of Bologna, where he teaches Analog
  Electronics, Applied Analog Electronics and Sensing to students of
  Electronic Engineering and Aerospace Engineering. He is also a faculty
  member of the Advanced Research Center on Electronic Systems (ARCES) at
  the University of Bologna. In 2008, 2009, 2011 he was a visiting
  researcher at the University of Washington in Seattle for short
  periods. His current research interests include nonlinear signal
  processing, internally nonlinear, externally linear networks, chaotic
  maps, delta-sigma modulation, testing of analog circuits, and random
  number generation. Sergio Callegari has authored or co-authored more
  than 70 papers in international conferences, journals and scientific
  books, as well as 4 national patents. In 2004 he was co-recipient of the
  IEEE Circuit and Systems Society Darlington Award, for the best paper
  appeared in the IEEE Transactions on Circuits and Systems in the
  previous biennium.  He served as an Associate Editor for the IEEE
  Transactions on Circuits and Systems --- Part II during 2006--2007 and
  as an Associate Editor for the IEEE Transactions on Circuits and Systems
  --- Part I during 2008--2009. He is a member of the Technical Committee
  on Nonlinear Circuits and Systems and of the Technical Committee on
  Education and Outreach of the IEEE CAS Society.  He also served in the
  Organization Committee and as Publication Co-Chair at NOLTA 2006
  (Bologna), as a member of the Organization Committee of Eurodoc 2006
  (Bologna), as a Special-Session Co-Chair of NOLTA 2010 (Krakow) and as a
  Chair for the nonlinear circuits track at ICECS 2012.  In 2005--2007, he
  has been a member of the board of the Italian Society of Doctoral
  Candidates and Ph.D.\@ Graduates (ADI) and currently he is in the board
  of AIR, a national society for the promotion of scientific research.
\end{IEEEbiography}
\begin{IEEEbiography}[{%
    \includegraphics[width=1in,height=1.25in,clip,keepaspectratio]{%
      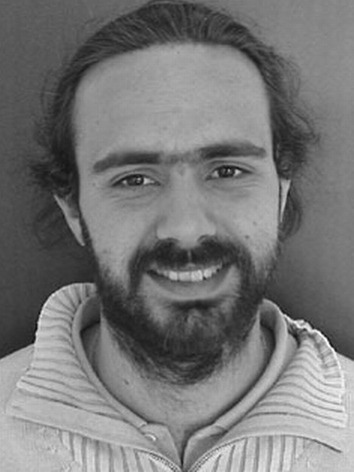}}]{%
    Federico Bizzarri}%
  (M12) was born in Genoa, Italy, in 1974.  He received the Laurea (M.Sc.)
  five-year degree (summa cum laude) in electronic engineering and the
  Ph.D. degree in electrical engineering from the University of Genoa,
  Genoa, Italy, in 1998 and 2001, respectively.  Since June 2010 he has
  been a temporary research contract Assistant rofessor at the Electronic
  and Information Department of the Politecnico di Milano, Milan, Italy.
  In 2000 he was a visitor to EPFL, Lausanne, Switzerland.  From 2002 to
  2008 he had been a post-doctoral research assistant in the Biophysical
  and Electronic Engineering Department of the University of Genova,
  Italy. In 2009 he was a post-doctoral research assistant in the ARCES
  Research Center of the University of Bologna, Italy.  His main research
  interests are in the area of nonlinear circuits, with emphasis on
  chaotic dynamics and bifurcation theory, circuit models of nonlinear
  systems, image processing, circuit theory and simulation.  He is the
  author or coauthor of about 50 scientific papers, more than an half of
  which have been published in international journals.
\end{IEEEbiography}
\vfill





\end{document}